\documentclass[%
 reprint,
superscriptaddress,
 amsmath,amssymb,
 aps,
]{revtex4-2}

\usepackage{graphicx}
\usepackage{dcolumn}
\usepackage{bm}
\usepackage{hyperref}

\makeatletter
\newsavebox{\@brx}
\newcommand{\llangle}[1][]{\savebox{\@brx}{\(\m@th{#1\langle}\)}%
  \mathopen{\copy\@brx\kern-0.5\wd\@brx\usebox{\@brx}}}
\newcommand{\rrangle}[1][]{\savebox{\@brx}{\(\m@th{#1\rangle}\)}%
  \mathclose{\copy\@brx\kern-0.5\wd\@brx\usebox{\@brx}}}
\makeatother

\begin{document}

\preprint{APS/123-QED}

\title{Superconductivity of repulsive spinless fermions with sublattice potentials}

\author{Yuchi He}
\email{yuchi.he@rwth-aachen.de}
\affiliation{Institut f{\"u}r Theorie der Statistischen Physik, RWTH Aachen University and
  JARA---Fundamentals of Future Information Technology, 52056 Aachen, Germany}
\author{Kang Yang}
\email{kang.yang@fu-berlin.de}
\affiliation{Department of Physics, Stockholm University, AlbaNova University Center, 106 91 Stockholm, Sweden}
\author{Jonas B.~Profe}
\email{profe@itp.uni-frankfurt.de}
\affiliation{Institut f{\"u}r Theorie der Statistischen Physik, RWTH Aachen University and
  JARA---Fundamentals of Future Information Technology, 52056 Aachen, Germany}
\author{Emil J.~Bergholtz}
\email{emil.bergholtz@fysik.su.se}
\affiliation{Department of Physics, Stockholm University, AlbaNova University Center, 106 91 Stockholm, Sweden}
\author{Dante M.~Kennes}
\email{dante.kennes@mpsd.mpg.de}
\affiliation{Institut f{\"u}r Theorie der Statistischen Physik, RWTH Aachen University and
  JARA---Fundamentals of Future Information Technology, 52056 Aachen, Germany}
 \affiliation{Max Planck Institute for the Structure and Dynamics of Matter, Center for Free Electron Laser Science, 22761 Hamburg, Germany}

\date{\today}
\begin{abstract}
We explore unconventional superconductivity of repulsive spinless fermions on square and honeycomb lattices with staggered sublattice potentials.  
The two lattices can exhibit staggered $d$-wave and $f$-wave pairing, respectively, at low doping stemming from an  effective two-valley band structure. At higher doping, in particular, the square lattice displays a much richer phase diagram including topological $p+ip$ superconductivity which is induced by a qualitatively different mechanism compared to the $d$-wave pairing. We illuminate this from several complementary perspectives: We analytically perform sublattice projection to analyze the effective continuum low-energy description and we numerically calculate the binding energies for pair and larger bound states for few-body doping near half filling. Furthermore, for finite doping, we present phase diagrams based on extensive functional renormalization group and and density matrix
renormalization group calculations. 
\end{abstract}

\maketitle

There have been substantial efforts~\cite{PhysRevLett.15.524,LeeNagaosaWen,science.1200181,Stewart} to understand superconductivity mechanisms beyond the conventional phonon-mediated~\cite{BCS} electron-electron attraction. In one category of mechanisms, 
bare repulsive electron-electron interaction becomes effectively attractive due to virtual processes after projections to the sublattice or bands
~\cite{PhysRev.134.A1416,PhysRevB.92.224514,dong2021activating}. Recently, exact results for an effective attraction have been obtained for fermionic honeycomb lattice models with a large staggered sublattice potential~\cite{sciadv,crepel2021spintriplet,crepel2021unconventional}. This mechanism can be essentially captured by a minimal model of spinless fermions~\cite{sciadv}, of which the low-energy physics projected to one sublattice shows effective attraction. Such a mechanism has been argued to be relevant for triplet pairing in materials~\cite{crepel2021spintriplet,science.abb9860,science.aar4642, science.aar4426}. 

In this Letter, we study the pairing of spinless fermions on the square lattice in addition to the honeycomb lattice model studied in Ref.~\cite{sciadv}. Studying a different lattice can shed light on the relevance of the proposed pairing mechanism to layered materials, in which different lattice structures can be realized~\cite{zhang2013}. Considering a different lattice contributes to further understanding the ingredients of the sublattice projection mechanism for superconductivity---and, as we show, reveals qualitatively different possibilities. The effective theory from a sublattice projection depends on the coordination number of the lattices; lattice symmetry is crucial for the realization of different types of unconventional superconductivity~\cite{RevModPhys.63.239,PhysRevB.81.024504,PhysRevB.98.045142,PhysRevB.104.155116,Gneist2022}.

\begin{figure}
    \centering
    \includegraphics[width=\columnwidth]{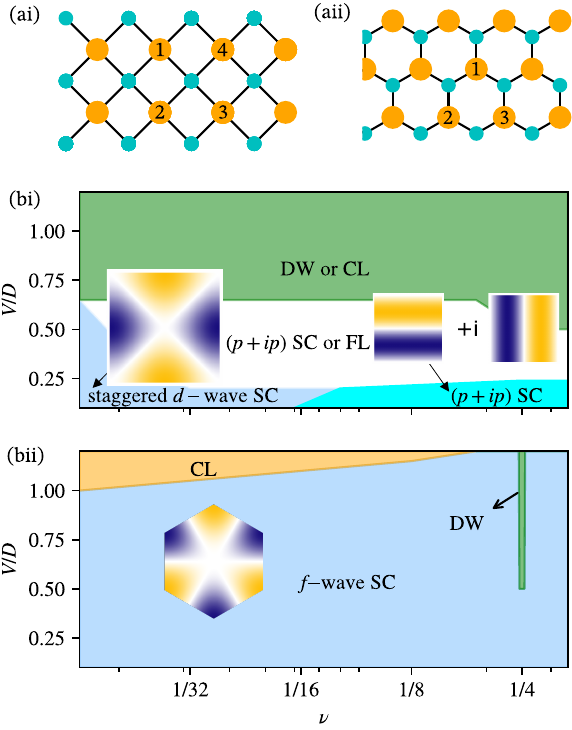}
    \caption{\textbf{Lattice structure and schematic phase diagrams.} (a) Lattice structure for (ai) square and (aii) honeycomb lattices; sublattices are marked in orange and cyan.  (b) Phase diagram for (bi) the square lattice with $t'=\lambda_0$  and (bii) the honeycomb lattice  with $t'=0$, inferred from FRG (weak coupling) and DMRG (strong coupling). For superconducting (SC) phases, we plot the momentum dependence of the susceptibility from FRG. For the $p+ip$ SC phase, a degenerate pair of dominant eigenvectors is found in FRG, and a mean-field analysis indicates the linear combination $p+ip$ is favored.  Here DW denotes the density-wave phase, CL denotes a phase separation via the collapse of electrons on the $B$ sublattice, and FL denotes Fermi liquids. For the uncertainty in the DMRG data interpretation, see the discussion in the main text.}
    \label{fig:pd}
\end{figure}

The overall result is summarized in Fig.~\ref{fig:pd}. The quantum phases are inferred through infinite density matrix renormalization group (DMRG)~\cite{White,mcculloch2008infinite} data for strong coupling combined with functional renormalization group (FRG)~\cite{RevModPhys.84.299} data at weak coupling.
Superconducting phases are found in a wide range of interaction parameters in the honeycomb model while its regime is limited to smaller interactions for the square model. Compared to a previous study \cite{sciadv}, a significant difference is that there are two superconducting phases on the square lattice, the staggered $d$-wave and the $p+ip$ topological phases, in contrast to the sole $f$-wave pairing on the honeycomb lattice. The $d$-wave pairing on the square lattice shares the same origin as the $f$-wave pairing on the honeycomb lattice in the sense of inter-valley pairing.
The Cooper pair arises from an inter-valley attraction revealed by sublattice projection.
This  requires a next-nearest-neighbor hopping $t'$ to 
realize a two-valley band structure for the square lattice. 
Upon increasing doping, 
we observe a transition from staggered $d$ to a topological $p+ip$~\cite{RG} superconductor. 
With zero momentum, $p+ip$ no longer results from the inter-valley attraction. It does not require the next-nearest-neighbor hopping.
Moreover, at stronger interactions, we find evidence for a transition from superconductivity to inhomogeneous states.

\emph{Model and low-energy description.--- }
We use the square lattice [Fig.~\ref{fig:pd}(ai)] as an example while the honeycomb model [Fig.~\ref{fig:pd}(aii)] can be found in Ref.~\cite{sciadv} with the same form of Hamiltonian. The Hamiltonian is taken as

\begin{align}
H=&\sum_{\langle i,j \rangle} \left[-\left(t c^{\dagger}_{i}c_{j}+\mathrm{H.c.}\right)+Vn_in_j\right]\nonumber\\&-\sum_{\llangle i,j \rrangle} \left(t'c^{\dagger}_{i}c_{j}+\mathrm{H.c.}\right) +\sum_{i \in B}Dn_i,\label{full model}  
\end{align}
where $c_i$, ($c_i^{\dagger}$) is the fermionic annihilation (creation)  operator on site $i$, and $n_i=c^{\dagger}_ic_i$. The symbols $\langle i,j \rangle$ and $\llangle i,j \rrangle$ denote nearest neighbors and next-nearest neighbors, respectively. We limit our attention to repulsive interaction $V>0$ and sublattice potential $D\gg |t|>0$ on the sublattice $B$. At half filling and large $D$, the  ground state is expected to have the $A$ sublattice fully filled and the $B$ sublattice unfilled. 
When $t'=0$, the Hamiltonian exhibits an explicit symmetry of particle-hole transformation $c_A^\dagger\to c_A$ and $c_B^\dagger\to -c_B$ combined with spatial inversion that interchanges the sublattices. When $t' \neq0$, the combined particle-hole transformation equivalently changes the sign of $t'$. In this work, we will only introduce $t'\ne 0$ on the square lattice while $t'=0$ on the honeycomb lattice, which is motivated by the discussion below.

We focus on electron doping the system near half filling, where the low-energy physics is controlled by those extra electrons on the $B$ lattice. The effective model is derived by a Schrieffer-Wolff transformation~\cite{MacDonald, Chernyshev} (for details see the Supplemental Material~\cite{suppm}). Up to the second order of $t$, this effective Hamiltonian contains terms of hopping, correlated hopping, and interactions: $H_{\textrm{eff}}=H_{\textrm{hopping}}+H_{\textrm{ch}}+H_{U}$.

Different parts of the Hamiltonian are introduced as follows (for the details of the coefficients see \cite{suppm}): $H_{\textrm{hopping}}$ contains nearest neighbors $\langle ij \rangle$
and next-nearest neighbors $\llangle ij \rrangle$ terms
for the sublattice $B$:
\begin{align}\label{full modelkine}
     H_{\textrm{hopping}}=&\left[\sum_{\langle ij \rangle} t_B c^\dagger_{i,B}c_{j,B}+\sum_{\llangle ij \rrangle} t'_B c^\dagger_{i,B}c_{j,B}\right]+\mathrm{H.c.},
\end{align}
where $t_B=2\lambda_0-t'$ and $t'_B=\lambda_0$ with $\lambda_0=t^2/(D+2V)$. For most of our calculations, we will either fix $t'=0$ or $t'=\lambda_0$.
 
The correlated hopping also includes two terms
\begin{align}
     H_{\textrm{ch}}=\sum_{ijk \in \square}   \lambda_1 c^\dagger_{i,B}c_{j,B}n_{k,B}+\sum_{ijkl \in \square}\frac{\lambda_2}{2}c^\dagger_{i,B}c_{j,B}n_{k,B}n_{l,B} \label{eq_swHK}
\end{align}
The combinations $ijk$ and $ijkl$ are summed over all possible ordered vertices of plaquettes in sublattice $B$, e.g., 1,2,3 and 4 in Fig.~\ref{fig:pd}(ai). Finally, there are two-, three-, and four-body density interactions,
\begin{align}
     H_{U}&=\sum_{\langle ij \rangle} 2U_2n_{i,B}n_{j,B}+\sum_{\llangle ij \rrangle} U_2 n_{i,B}n_{j,B}+ \nonumber \\
     &\sum_{[ijk] \in \square} U_3n_{i,B}n_{j,B}n_{k,B}+\sum_{[ijkl] \in \square}U_4n_{i,B}n_{j,B}n_{k,B}n_{l,B},\label{eq_swHU}
\end{align}
The combinations $[ijk]$ and $[ijkl]$ are summed over all possible unordered vertices of plaquettes in sublattice $B$. The four-body interaction $U_4$ remains repulsive in the full parameter region, while other interaction terms turn from repulsion to attraction when increasing across $V/D=1$.

\begin{figure}
    \centering
    \includegraphics[width=\columnwidth]{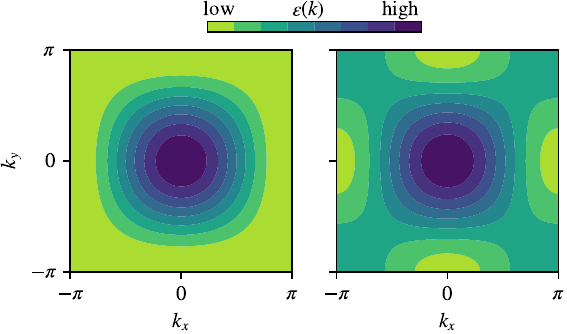}
    \caption{\textbf{Band structures of the square lattice's effective model's kinetic part.} at (a) $t'=0$ and (b)  $t'=\lambda_0$. The right band structure has two valleys at $(0, \pm \pi)$ and $(\pm \pi,0)$, respectively. The two-valley structure is absent for $t'=0$ [in (a)]. 
    }
    \label{fig:bandstructure}
\end{figure}

The dispersion of the kinetic part $H_{\textrm{hopping}}$ depends on the next-nearest-neighbor hopping $t'$. At $t'=0$ [shown in Fig.~\ref{fig:bandstructure} (a)], the band minimum is located along the boundary of the Brillouin zone. The Fermi surface is connected and has an approximate rotation symmetry. By tuning $t'$ such that $|t'_B/t_B|>0.5$, two band minima appear at $(0, \pm \pi)$ and $(\pm \pi,0)$, respectively, where the unit of the wave vectors is $1/a$. The low-energy physics is then controlled by these two valleys which are interchanged under a $\pi/2$ rotation. When tuning to higher doping, the Fermi surface includes the Van Hove singularities. They are located at $(q,\pm q)$
with $q=\pm \arccos (-t_B/(2t'_B))$. The two-valley low-energy physics is replaced by the one exhibiting new instabilities driven by the larger density of states. We remark that introducing $t'$ on the honeycomb lattice only brings an overall factor to the band dispersion.

\emph{Two-valley continuum theory of the square lattice model. --- }
To construct the continuum theory in the case with two valleys, the degrees of freedom for doped electrons can be decomposed into two valleys:
{$c_j=\sum_{\sigma}a\exp[i\mathbf  K_\sigma \cdot \mathbf r_j]\psi_{\sigma}(\mathbf r_j)$} with $\mathbf  K_+=(0,\pi)$ and $\mathbf  K_-=(\pi,0)$, where the fields $\psi_{\sigma}(\mathbf r)$ vary slowly at the scale of $a$, the minimal distance between two $B$-sublattice sites.

At low doping, we ignore the three- and four-body interactions in $H_U$. The continuum Hamiltonian includes a kinetic part with anisotropic masses $\sum_\sigma \psi^\dagger_\sigma\partial_x^2\psi_\sigma/(2 m_\sigma^{xx})+\psi^\dagger_\sigma\partial_y^2\psi_\sigma/(2 m_\sigma^{yy})$ at two valleys and a two-body interaction term. 
There are two contributions to the two-body interaction in the continuum limit, the correlated hopping terms in Eq.~\eqref{eq_swHK} and the two-body repulsion terms Eq.~\eqref{eq_swHU}. In the long-wavelength limit, the interaction can be written as
\begin{align}
\int d^2\mathbf{r} g\psi^{\dagger}_{+}(\bold{r})\psi_{+}(\bold{r})\psi^{\dagger}_{-}(\bold{r})\psi_{-}(\bold{r}),
\end{align}
where $g=(16U_2-32\lambda_1)a^2=16a^2[-4t^2/(D+3V)+8t^2(D+2V)-4t^2/(D+3V)]<0$, indicating two-particle ground states are always inter-valley bound states. A possible low-doping superconducting (SC) state arises from a condensate of inter-valley pairing $\langle  \psi_{+}(\bold{r})\psi_{-}(\bold{r} )\rangle \neq 0$. In terms of microscopic fields, we find a total momentum $(\pi,\pi)$, $d_{x^2-y^2}$ pairing with an order parameter $\langle c_i c_j \rangle =[(-1)^{i_x+j_y}-(-1)^{i_y+j_x}]\Delta(i-j)$, where $\Delta$ is odd under a $\pi/2$ rotation.  While the pair has non-zero total momentum, the above reasoning for the pairing is the same as that for $f$-wave SC of low-doping honeycomb model~\cite{crepel2021unconventional}. For finite doping, realizing pairing with $(\pi,\pi)$ center-of-mass momentum is frustrated by the shape of the Fermi surfaces. This could lead to a transition to incommensurate (not observed) or other SC phases. Inferring the possible SC at finite doping from the bare Hamiltonian of the projected model is no longer simple. The complication comes from the interactions projected on the Fermi surface. Nevertheless, we can show that for the intra-valley interaction, the correlated hopping in the projected model can induce bare attractive interaction term  between  pairs of fermion modes on the Fermi surface with zero net momentum, for details see the  the Supplemental Material~\cite{suppm}. Thereby the possibility of intra-valley pairing, likely $p$-wave pairing, is suggested. We will later discuss the role of Van Hove singularity for SC, which is independent of the role of projected interactions.

\begin{figure*}
    \centering
    \includegraphics[width=\linewidth]{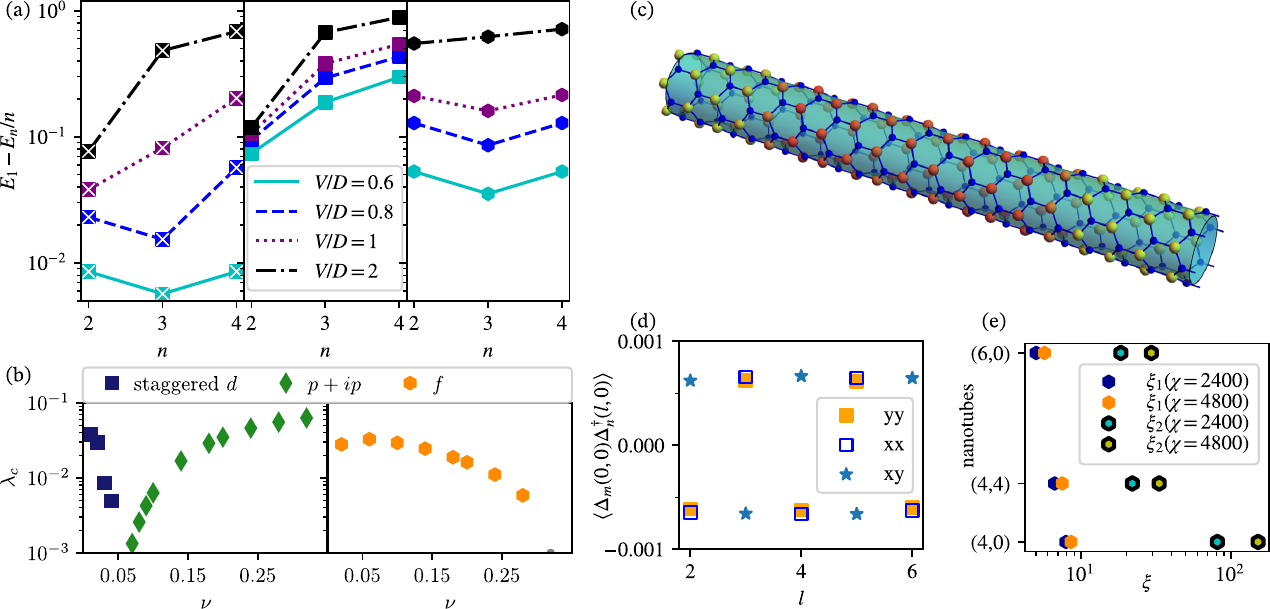}
    \caption{\textbf{Results from exact diagonalization (ED) of the effective model for few-particle doping, with FRG and DMRG of the full model for finite doping.} (a) Energy per particle of the $n$-particle ground states; the unit is $t^2/D$. Calculations are performed for the effective models using ED with finite-size extrapolation. From left to right are the square lattice model with $t'=\lambda_0$ and the honeycomb lattice $t'=0$. Finite $E_1-E_2/2$ indicates the existence of two-particle bound states. Some paired phase near the dilute limit is indicated by the energy per particle $E_{1}-E_{n}/n$ being a constant for every positive even $n$ and larger than the values for odd $n$. On the other hand, if there is some $n$, with $E_{1}-E_{n}/n$ greater than $E_1-E_2/2$, larger bound states are favored. (b) FRG predicted phases and energy scales $\lambda_c$ in units of $t^2/D$. Shown on the left is the square lattice model with $t'=\lambda_0$ and on the right is the honeycomb lattice. We choose $V = 2t$ as the interaction. (c) Cylinder geometry [(6, 0) honeycomb nanotube] and a density plot of an inhomogeneous density profile, indicating phase separation. The densities on $B$ sublattice are represented by colors with green for small density and red for larger density. The parameters for the density plot are $\frac{1}{8}$ doping, $D=10, V=15$, and unit cell size 32; only part of the unit cell is plotted. (d) Pair correlation functions for square lattice at $\frac{1}{64}$ doping, $V=6, D=10$ and a tangential direction size eight unit cells. For the definitions of pair operators see the text; the subscripts $m,n$ can be $x,y$. (e) Correlation length evidence for pairing on the infinite cylinder geometry of a honeycomb lattice. The indices for nanotubes are standard notations for their sizes and shapes. The $\xi_1$ ($\xi_2$) is the single-particle (pair) correlation length. For a larger bond dimension $\chi$ used in the iDMRG algorithm, a tighter lower bound of $\xi$ is obtained. Here, a smaller $\xi_2(\chi)$ for a larger nanotube is an artifact of the underestimation becomes more severe for larger systems for fixed $\chi$. The other parameters are $\frac{1}{8}$ doping and $D=V=5$. }
    \label{fig:energyperparticle}
\end{figure*}

\emph{Binding energies for few-particle doping. --- }
Next, we show our numerical results of pair and larger bound states formation in the dilute doping limit.  Binding energies can be deduced from the difference between one-particle doping energy and energy per particle of n-particle doping; the data for the effective model ($D/t=\infty$) are plotted in Fig.~\ref{fig:energyperparticle}. (Our data for $D/t=5,10$ can be found in Ref.~\cite{suppm}.)  From the data, we can infer that at $D/t=\infty$, there can be a stable dilute pairing phase for the honeycomb lattice  with $V/D \lessapprox 1$.  The pairing phase is not favored for the square lattice with $t'=0$,  but it can exist with $t'>0$. For $t'=\lambda_{0}$, the condition for pairing phase is $V/D \lessapprox 0.6$.

We also determine the momenta of the few-particle ground states. The momentum of a pair for the square lattice with $t'=\lambda_0$ and the honeycomb lattice, are respectively $(\pi,\pi)$ and $(0,0)$. 
Recall that two valleys of the honeycomb lattice are located at $\pm \bold{K}$ (standard notation~\cite{Castro_Neto_2009}), and those of the square lattice model are located at $(\pi,0)$ or $(0,\pi)$. 
This along with finite pair binding energy results indicates an inter-valley pairing mechanism and explains the absence of it in the case of $t'=0$ with the absence of valleys. The two-valley structure allows stable pairing, for which a sufficient attraction between fermions in different valleys exists but no attractions sufficient for larger bound states.
The latter condition can be usually met with weak coupling as the intra-valley coupling is less relevant in the dilute doping limit.

\emph{Numerical study of the quantum phases at finite doping.--- }
The above few-body and continuum theory results provide an indication of superconductivity at low doping and its instability for large interactions. In the following, we apply DMRG and FRG to infer the quantum phases of the full models with $D=10$ and $t'=\lambda_0$ (square) and $t'=0$ (honeycomb) at finite fermion doping from weak to strong coupling (details see the Supplemental Material~\cite{suppm}); the results are summarized in Fig.~\ref{fig:pd}. The $d_{x^2-y^2}$- and $f$-wave superconductivity of square and honeycomb lattice expected at dilute doping are observed by both methods. Upon increasing doping of the square lattice by approximately $0.1$, our FRG calculation indicates a $p+ip$ superconducting phase.  In the honeycomb lattice model, the $f$-wave superconductivity persists for higher doping, corroborating the main claim of Refs.~\cite{sciadv,crepel2021unconventional}; but our DMRG data suggest the absence of superconductivity at the Van Hove singularity $\nu=\frac{1}{4}$, in contrast to Refs.~\cite{sciadv, crepel2021unconventional}. Recall that near the Van Hove singularity the two-valley picture breaks down.

In the weak coupling regime ($V/D \lessapprox 0.3$), we perform FRG calculations~\cite{RevModPhys.84.299,Dupuis_2021} at the one-loop level. 
We only include the static self-energy and the static two-particle interaction. More particle processes are only included as virtual processes in the two-particle vertex. The inclusion of the static self-energy has been shown to cover already the relevant physics in one-dimensional (1D) systems~\cite{Markhof_2018} and can be argued to cover the relevant physics more generally by power counting arguments~\cite{RevModPhys.84.299}. The static self-energy incorporates possible further increases of the band gap and deformations of the Fermi surface. For our simulations, 
we use the unified truncated unity (dubbed $\mathrm{TU}^{2}$) approach~\cite{Profe_2022} merging real and momentum space, which has been demonstrated to fulfill in the FRG equivalence class~\cite{Beyer_2022}.  
We distinguish different phases in our FRG simulations by inspecting the eigenvectors corresponding to the largest eigenvalue of each diagrammatic channel. Each of these channels corresponds to a different type of instability and the symmetry of the eigenvector gives the symmetry of the ordering. The Fourier transformation of these eigenvectors at the $B$ sublattice are visualized as insets in Fig.~\ref{fig:pd}.
In the strong-coupling regime ($V/D \gtrapprox 0.5$), we use DMRG~\cite{White, mcculloch2008infinite} to obtain ground states on infinite cylinder geometries~\cite{CNT}. We consider cylinders with up to eight sites along the circumference. 
The counterpart of 2D superconductivity on the cylinder cannot retain long-range order because of the Mermin-Wagner theorem. However, the pair correlation is expected to be dominant over single-particle correlation.  In most common cases, the single-particle excitation of a quasi-1D system is fully gapped (see, e.g., Ref.~\cite{peng2022enhanced}); the single particle correlation length $\xi_1$ is finite, while the pair correlation length $\xi_2$ can diverge.
Thus, observation of estimated $\xi_2 \gg \xi_1$ serves as evidence for such pairing. 
The DMRG estimation~\cite{Zauner_2015,PhysRevB.104.195126,pollmann2009theory} usually sets lower bounds for correlation lengths, which become tighter with an increasing number of variational parameters characterized by bond dimension $\chi$~\cite{pollmann2009theory}.

For the square lattice, we find the predicted $d_{x^2-y^2}$-wave superconductor at low doping within our FRG simulations with $t'=\lambda_0$. However, the critical energy scale drops rapidly upon increasing doping and at higher doping ($\nu \approx 0.1$) we observe a transition to a $p_{x}+ip_{y}$ topological superconductor with Chern number~\cite{Fukui_2005} $\mathcal{C} = 2$. The low-doping phase is expected from the above bare interaction analysis. Besides the bare interaction terms revealed above, we speculate a density-fluctuation mediated mechanism~\cite{PhysRevB.35.6694} can be crucial for stabilizing the higher-doping SC.  One observation is a clear increase of SC energy scale at higher doping closer to the density state maxima [Fig.~\ref{fig:energyperparticle}(b)].
The transition between the phases seems to be driven by a change of weight within the particle-particle loop, whereupon doping the $d_{x^2-y^2}$ eigenvector will be increasingly suppressed while the $p_{x}/p_{y}$ pair will increase in strength. At stronger interactions ($V/D \approx 0.25$) our FRG breaks down, manifested as a linear ramp-up of the density-density interaction. This ramp-up marks the breakdown of the perturbative regime and hence FRG cannot be used to examine the phases.  We additionally study the case $t'=0$ for which the two valleys are absent, the ramp-up problem exists at low doping even for weak interaction, and a $\mathcal C=1$ $p+ip$ is observed for some higher doping~\cite{suppm}.
From the DMRG data ($V/D \gtrapprox 0.5$ and $t'=\lambda_0$), a finite  single-particle correlation length $\xi_1$ is only consistently found in low doping approximately $\frac{1}{64}$ and intermediate interaction; in this case, the pair correlation shows a dominant oscillatory part, supporting the staggered $d_{x^2-y^2}$ pairing [Fig.~\ref{fig:energyperparticle}(d)]. We study the geometry with the axial direction along the shortest lattice unit vector (e.g., that connecting nodes 1 and 2 in Fig.~\ref{fig:pd}(a1) right). We consider two-point correlations between $\Delta_x(i_x,i_y)= c_{i_x,i_y}c_{i_x+1,i_y}$, $\Delta_y(i_x,i_y)= c_{i_x,i_y}c_{i_x,i_y+1}$ and their Hermitian conjugation. Only the sites on $B$ sublattice are considered. We observe that the signs of $\langle \Delta_x(0,0) \Delta^{\dagger}_x(l,0) \rangle$ and $\langle \Delta_y(0,0) \Delta^{\dagger}_y(l,0) \rangle$ oscillate in $l$; we also observe that the sign of $\langle \Delta_x(0,0) \Delta^{\dagger}_y(l,0) \rangle$ is opposite to the previous two for a given $l$. For higher doping, no evidence of convergent $\xi_1$ is found and no evidence for time-reversal symmetry breaking is found for the implemented larger bond dimensions. While these can be features of a quasi-1-D analog of a Fermi liquid (FL), topological  $p+ip$ pairing cannot be excluded. The particle-number-conserved 1D analog of the topological $p$-wave state has been suggested to be adiabatically connected to an FL~\cite{KSH}; a deeper understanding of the quasi-1-D analog of $p+ip$ is needed to better interpret the data for the $p+ip$ SC or FL region of Fig.~\ref{fig:pd}(b1). The region for large $V/D$ denoted by DW in Fig.~\ref{fig:pd}, is characterized by inhomogeneous densities within the implemented bond dimensions. The 2D phases are speculated to be charge density waves at sufficient commensurate doping; other doping could be Fermi liquids or phases separated by Maxwell construction. The density-wave patterns are difficult to determine as they may only fit on larger cylinders than those studied.  

For the honeycomb lattice, we observe $f$-wave superconductivity in FRG for a broad range of doping, which exceeds Van Hove doping $\frac{1}{4}$. The range is slightly smaller than the random-phase approximation result Ref.~\cite{crepel2021unconventional}.
Similar to the square lattice, there is also a ramp-up refraining FRG prediction at stronger coupling. Our DMRG for stronger coupling  shows a broad range for pairing with a single-particle gap. This is observed for all geometries we studied (e.g., Fig.~\ref{fig:energyperparticle}(e)),  including the zigzag and armchair nanotube geometries, denoted by $(n,0)$ and $(n,n)$ [standard notation~\cite{CNT}] respectively where $n$ characterizes the circumference. However, right at the Van Hove doping $\nu=\frac{1}{4}$, most cylinder setups including the largest, point to insulating states~\cite{suppm}. This feature indicates a possible mechanism of density-fluctuation-induced SC which can accompany a density-wave phase at a commensurate filling~\cite{PhysRevB.98.045142}. This independent mechanism for $f$-wave SC provides an explanation why the phase extends to higher doping compared to the previous estimation of bare interaction~\cite{crepel2021unconventional}. However, the SC energy scale is not largely enhanced closer to the density of states maxima [Fig.~\ref{fig:energyperparticle}(b)], in contrast to the square lattice. This point is further supported by the high doping state still being an $f$-wave superconductor~\cite{PhysRevB.98.045142}, such that no competition between the mechanisms is realized. The CL indicates the collapse of fermions leaving part of the system with vanishing occupancy on the $B$ sublattice; collapses are usually observed for models with strong attractive interactions~\cite{emtll,PhysRevResearch.3.013114, PhysRevB.104.155116}. 
The observation is that the fermions on lattice  B always concentrate on part of the unit cell [see Fig.~\ref{fig:energyperparticle}(c)] when increasing iMPS unit cell size.

\emph{Discussion.--- }
We examined fermion pairing  driven by repulsive interaction and a strong sublattice potential for square lattices and honeycomb lattices.  The honeycomb lattice is confirmed to show $f$-wave pairing, which can be interpreted as inter-valley pairing.  The square lattice's counterpart of inter-valley pairing is found to  give a low-doping $d$-wave superconductivity with $(\pi,\pi)$ total momentum. Upon increasing doping, a $p+ip$ topological superconductivity is found.   Because of the role and existence condition of valleys, the square lattice model with next-nearest-neighbor hopping can exhibit an asymmetry for electron and hole doping. 
As an outlook, one may also include spin degrees of freedom~\cite{Xiao_2016,crepel2021spintriplet} and more types of interactions and hoppings, which serve as extensions of ionic Hubbard models~\cite{PhysRevLett.115.115303,PhysRevB.76.085112,JPSJ,PhysRevB.99.155127,10.21468/SciPostPhysCore.4.2.009}. This may have implications for real materials and provide the possibility of the sought-after $p+ip$ superconductivity with topological order.

\emph{Acknowledgments.--- }We thank Johan Carlström, Biao Huang, Ciar\'an Hickey, Dominik Kiese, Lennart Klebl, Lukas Weber, Stefan Wessel, and Fan Yang for discussions. The QuSpin~\cite{10.21468/SciPostPhys.7.2.020} and TeNPy~\cite{10.21468/SciPostPhysLectNotes.5} packages were used for the numerical studies. 
The authors gratefully acknowledge  the computing time granted by the Max Planck Computing and Data Facility.
The authors gratefully acknowledge the computing time granted by the JARA Vergabegremium and provided on the JARA Partition part of the supercomputer JURECA~\cite{JUWELS} at Forschungszentrum Jülich.
Y.H., J.B.H., and D.M.K. were supported by the Deutsche Forschungsgemeinschaft (German Research Foundation) under Grant No. RTG 1995, within the Priority Program
SPP 2244 ``2DMP'' and under Germany's Excellence Strategy- Cluster of
Excellence Matter and Light for Quantum Computing (ML4Q) Grant No. EXC 2004/1 -
390534769.
K.Y. and E.J.B. were supported by the Swedish Research Council (VR, 2018-00313), and the Wallenberg Academy Fellows program of the Knut and Alice Wallenberg Foundation (Grant No. 2018.0460).

\bibliography{bib}

\begin{thebibliography}{51}%
\makeatletter
\providecommand \@ifxundefined [1]{%
 \@ifx{#1\undefined}
}%
\providecommand \@ifnum [1]{%
 \ifnum #1\expandafter \@firstoftwo
 \else \expandafter \@secondoftwo
 \fi
}%
\providecommand \@ifx [1]{%
 \ifx #1\expandafter \@firstoftwo
 \else \expandafter \@secondoftwo
 \fi
}%
\providecommand \natexlab [1]{#1}%
\providecommand \enquote  [1]{``#1''}%
\providecommand \bibnamefont  [1]{#1}%
\providecommand \bibfnamefont [1]{#1}%
\providecommand \citenamefont [1]{#1}%
\providecommand \href@noop [0]{\@secondoftwo}%
\providecommand \href [0]{\begingroup \@sanitize@url \@href}%
\providecommand \@href[1]{\@@startlink{#1}\@@href}%
\providecommand \@@href[1]{\endgroup#1\@@endlink}%
\providecommand \@sanitize@url [0]{\catcode `\\12\catcode `\$12\catcode
  `\&12\catcode `\#12\catcode `\^12\catcode `\_12\catcode `\%12\relax}%
\providecommand \@@startlink[1]{}%
\providecommand \@@endlink[0]{}%
\providecommand \url  [0]{\begingroup\@sanitize@url \@url }%
\providecommand \@url [1]{\endgroup\@href {#1}{\urlprefix }}%
\providecommand \urlprefix  [0]{URL }%
\providecommand \Eprint [0]{\href }%
\providecommand \doibase [0]{https://doi.org/}%
\providecommand \selectlanguage [0]{\@gobble}%
\providecommand \bibinfo  [0]{\@secondoftwo}%
\providecommand \bibfield  [0]{\@secondoftwo}%
\providecommand \translation [1]{[#1]}%
\providecommand \BibitemOpen [0]{}%
\providecommand \bibitemStop [0]{}%
\providecommand \bibitemNoStop [0]{.\EOS\space}%
\providecommand \EOS [0]{\spacefactor3000\relax}%
\providecommand \BibitemShut  [1]{\csname bibitem#1\endcsname}%
\let\auto@bib@innerbib\@empty
\bibitem [{\citenamefont {Kohn}\ and\ \citenamefont
  {Luttinger}(1965)}]{PhysRevLett.15.524}%
  \BibitemOpen
  \bibfield  {author} {\bibinfo {author} {\bibfnamefont {W.}~\bibnamefont
  {Kohn}}\ and\ \bibinfo {author} {\bibfnamefont {J.~M.}\ \bibnamefont
  {Luttinger}},\ }\bibfield  {title} {\bibinfo {title} {New mechanism for
  superconductivity},\ }\href {https://doi.org/10.1103/PhysRevLett.15.524}
  {\bibfield  {journal} {\bibinfo  {journal} {Phys. Rev. Lett.}\ }\textbf
  {\bibinfo {volume} {15}},\ \bibinfo {pages} {524} (\bibinfo {year}
  {1965})}\BibitemShut {NoStop}%
\bibitem [{\citenamefont {Lee}\ \emph {et~al.}(2006)\citenamefont {Lee},
  \citenamefont {Nagaosa},\ and\ \citenamefont {Wen}}]{LeeNagaosaWen}%
  \BibitemOpen
  \bibfield  {author} {\bibinfo {author} {\bibfnamefont {P.~A.}\ \bibnamefont
  {Lee}}, \bibinfo {author} {\bibfnamefont {N.}~\bibnamefont {Nagaosa}},\ and\
  \bibinfo {author} {\bibfnamefont {X.-G.}\ \bibnamefont {Wen}},\ }\bibfield
  {title} {\bibinfo {title} {{Doping a Mott insulator: Physics of
  high-temperature superconductivity}},\ }\href
  {https://doi.org/10.1103/RevModPhys.78.17} {\bibfield  {journal} {\bibinfo
  {journal} {Rev. Mod. Phys.}\ }\textbf {\bibinfo {volume} {78}},\ \bibinfo
  {pages} {17} (\bibinfo {year} {2006})}\BibitemShut {NoStop}%
\bibitem [{\citenamefont {Norman}(2011)}]{science.1200181}%
  \BibitemOpen
  \bibfield  {author} {\bibinfo {author} {\bibfnamefont {M.~R.}\ \bibnamefont
  {Norman}},\ }\bibfield  {title} {\bibinfo {title} {{The Challenge of
  Unconventional Superconductivity}},\ }\href
  {https://doi.org/10.1126/science.1200181} {\bibfield  {journal} {\bibinfo
  {journal} {Science}\ }\textbf {\bibinfo {volume} {332}},\ \bibinfo {pages}
  {196} (\bibinfo {year} {2011})}\BibitemShut {NoStop}%
\bibitem [{\citenamefont {Stewart}(2017)}]{Stewart}%
  \BibitemOpen
  \bibfield  {author} {\bibinfo {author} {\bibfnamefont {G.~R.}\ \bibnamefont
  {Stewart}},\ }\bibfield  {title} {\bibinfo {title} {{Unconventional
  superconductivity}},\ }\href {https://doi.org/10.1080/00018732.2017.1331615}
  {\bibfield  {journal} {\bibinfo  {journal} {Advances in Physics}\ }\textbf
  {\bibinfo {volume} {66}},\ \bibinfo {pages} {75} (\bibinfo {year}
  {2017})}\BibitemShut {NoStop}%
\bibitem [{\citenamefont {Bardeen}\ \emph {et~al.}(1957)\citenamefont
  {Bardeen}, \citenamefont {Cooper},\ and\ \citenamefont {Schrieffer}}]{BCS}%
  \BibitemOpen
  \bibfield  {author} {\bibinfo {author} {\bibfnamefont {J.}~\bibnamefont
  {Bardeen}}, \bibinfo {author} {\bibfnamefont {L.~N.}\ \bibnamefont
  {Cooper}},\ and\ \bibinfo {author} {\bibfnamefont {J.~R.}\ \bibnamefont
  {Schrieffer}},\ }\bibfield  {title} {\bibinfo {title} {{Theory of
  Superconductivity}},\ }\href {https://doi.org/10.1103/PhysRev.108.1175}
  {\bibfield  {journal} {\bibinfo  {journal} {Phys. Rev.}\ }\textbf {\bibinfo
  {volume} {108}},\ \bibinfo {pages} {1175} (\bibinfo {year}
  {1957})}\BibitemShut {NoStop}%
\bibitem [{\citenamefont {Little}(1964)}]{PhysRev.134.A1416}%
  \BibitemOpen
  \bibfield  {author} {\bibinfo {author} {\bibfnamefont {W.~A.}\ \bibnamefont
  {Little}},\ }\bibfield  {title} {\bibinfo {title} {{Possibility of
  Synthesizing an Organic Superconductor}},\ }\href
  {https://doi.org/10.1103/PhysRev.134.A1416} {\bibfield  {journal} {\bibinfo
  {journal} {Phys. Rev.}\ }\textbf {\bibinfo {volume} {134}},\ \bibinfo {pages}
  {A1416} (\bibinfo {year} {1964})}\BibitemShut {NoStop}%
\bibitem [{\citenamefont {Chen}\ \emph {et~al.}(2015)\citenamefont {Chen},
  \citenamefont {Maiti}, \citenamefont {Linscheid},\ and\ \citenamefont
  {Hirschfeld}}]{PhysRevB.92.224514}%
  \BibitemOpen
  \bibfield  {author} {\bibinfo {author} {\bibfnamefont {X.}~\bibnamefont
  {Chen}}, \bibinfo {author} {\bibfnamefont {S.}~\bibnamefont {Maiti}},
  \bibinfo {author} {\bibfnamefont {A.}~\bibnamefont {Linscheid}},\ and\
  \bibinfo {author} {\bibfnamefont {P.~J.}\ \bibnamefont {Hirschfeld}},\
  }\bibfield  {title} {\bibinfo {title} {{Electron pairing in the presence of
  incipient bands in iron-based superconductors}},\ }\href
  {https://doi.org/10.1103/PhysRevB.92.224514} {\bibfield  {journal} {\bibinfo
  {journal} {Phys. Rev. B}\ }\textbf {\bibinfo {volume} {92}},\ \bibinfo
  {pages} {224514} (\bibinfo {year} {2015})}\BibitemShut {NoStop}%
\bibitem [{\citenamefont {Dong}\ and\ \citenamefont
  {Levitov}(2021)}]{dong2021activating}%
  \BibitemOpen
  \bibfield  {author} {\bibinfo {author} {\bibfnamefont {Z.}~\bibnamefont
  {Dong}}\ and\ \bibinfo {author} {\bibfnamefont {L.}~\bibnamefont {Levitov}},\
  }\bibfield  {title} {\bibinfo {title} {{Activating superconductivity in a
  repulsive system by high-energy degrees of freedom}},\ }\href@noop {}
  {\bibfield  {journal} {\bibinfo  {journal} {arXiv preprint arXiv:2103.08767}\
  } (\bibinfo {year} {2021})}\BibitemShut {NoStop}%
\bibitem [{\citenamefont {Crépel}\ and\ \citenamefont {Fu}(2021)}]{sciadv}%
  \BibitemOpen
  \bibfield  {author} {\bibinfo {author} {\bibfnamefont {V.}~\bibnamefont
  {Crépel}}\ and\ \bibinfo {author} {\bibfnamefont {L.}~\bibnamefont {Fu}},\
  }\bibfield  {title} {\bibinfo {title} {{New mechanism and exact theory of
  superconductivity from strong repulsive interaction}},\ }\href
  {https://doi.org/10.1126/sciadv.abh2233} {\bibfield  {journal} {\bibinfo
  {journal} {Science Advances}\ }\textbf {\bibinfo {volume} {7}},\ \bibinfo
  {pages} {eabh2233} (\bibinfo {year} {2021})}\BibitemShut {NoStop}%
\bibitem [{\citenamefont {Crépel}\ and\ \citenamefont
  {Fu}(2022)}]{crepel2021spintriplet}%
  \BibitemOpen
  \bibfield  {author} {\bibinfo {author} {\bibfnamefont {V.}~\bibnamefont
  {Crépel}}\ and\ \bibinfo {author} {\bibfnamefont {L.}~\bibnamefont {Fu}},\
  }\bibfield  {title} {\bibinfo {title} {Spin-triplet superconductivity from
  excitonic effect in doped insulators},\ }\href
  {https://doi.org/10.1073/pnas.2117735119} {\bibfield  {journal} {\bibinfo
  {journal} {Proc. Natl. Acad. Sci. U.S.A.}\ }\textbf {\bibinfo {volume}
  {119}},\ \bibinfo {pages} {e2117735119} (\bibinfo {year} {2022})}\BibitemShut
  {NoStop}%
\bibitem [{\citenamefont {Cr\'epel}\ \emph {et~al.}(2022)\citenamefont
  {Cr\'epel}, \citenamefont {Cea}, \citenamefont {Fu},\ and\ \citenamefont
  {Guinea}}]{crepel2021unconventional}%
  \BibitemOpen
  \bibfield  {author} {\bibinfo {author} {\bibfnamefont {V.}~\bibnamefont
  {Cr\'epel}}, \bibinfo {author} {\bibfnamefont {T.}~\bibnamefont {Cea}},
  \bibinfo {author} {\bibfnamefont {L.}~\bibnamefont {Fu}},\ and\ \bibinfo
  {author} {\bibfnamefont {F.}~\bibnamefont {Guinea}},\ }\bibfield  {title}
  {\bibinfo {title} {{Unconventional superconductivity due to interband
  polarization}},\ }\href {https://doi.org/10.1103/PhysRevB.105.094506}
  {\bibfield  {journal} {\bibinfo  {journal} {Phys. Rev. B}\ }\textbf {\bibinfo
  {volume} {105}},\ \bibinfo {pages} {094506} (\bibinfo {year}
  {2022})}\BibitemShut {NoStop}%
\bibitem [{\citenamefont {Nakagawa}\ \emph {et~al.}(2021)\citenamefont
  {Nakagawa}, \citenamefont {Kasahara}, \citenamefont {Nomoto}, \citenamefont
  {Arita}, \citenamefont {Nojima},\ and\ \citenamefont
  {Iwasa}}]{science.abb9860}%
  \BibitemOpen
  \bibfield  {author} {\bibinfo {author} {\bibfnamefont {Y.}~\bibnamefont
  {Nakagawa}}, \bibinfo {author} {\bibfnamefont {Y.}~\bibnamefont {Kasahara}},
  \bibinfo {author} {\bibfnamefont {T.}~\bibnamefont {Nomoto}}, \bibinfo
  {author} {\bibfnamefont {R.}~\bibnamefont {Arita}}, \bibinfo {author}
  {\bibfnamefont {T.}~\bibnamefont {Nojima}},\ and\ \bibinfo {author}
  {\bibfnamefont {Y.}~\bibnamefont {Iwasa}},\ }\bibfield  {title} {\bibinfo
  {title} {{Gate-controlled BCS-BEC crossover in a two-dimensional
  superconductor}},\ }\href {https://doi.org/10.1126/science.abb9860}
  {\bibfield  {journal} {\bibinfo  {journal} {Science}\ }\textbf {\bibinfo
  {volume} {372}},\ \bibinfo {pages} {190} (\bibinfo {year}
  {2021})}\BibitemShut {NoStop}%
\bibitem [{\citenamefont {Fatemi}\ \emph {et~al.}(2018)\citenamefont {Fatemi},
  \citenamefont {Wu}, \citenamefont {Cao}, \citenamefont {Bretheau},
  \citenamefont {Gibson}, \citenamefont {Watanabe}, \citenamefont {Taniguchi},
  \citenamefont {Cava},\ and\ \citenamefont
  {Jarillo-Herrero}}]{science.aar4642}%
  \BibitemOpen
  \bibfield  {author} {\bibinfo {author} {\bibfnamefont {V.}~\bibnamefont
  {Fatemi}}, \bibinfo {author} {\bibfnamefont {S.}~\bibnamefont {Wu}}, \bibinfo
  {author} {\bibfnamefont {Y.}~\bibnamefont {Cao}}, \bibinfo {author}
  {\bibfnamefont {L.}~\bibnamefont {Bretheau}}, \bibinfo {author}
  {\bibfnamefont {Q.~D.}\ \bibnamefont {Gibson}}, \bibinfo {author}
  {\bibfnamefont {K.}~\bibnamefont {Watanabe}}, \bibinfo {author}
  {\bibfnamefont {T.}~\bibnamefont {Taniguchi}}, \bibinfo {author}
  {\bibfnamefont {R.~J.}\ \bibnamefont {Cava}},\ and\ \bibinfo {author}
  {\bibfnamefont {P.}~\bibnamefont {Jarillo-Herrero}},\ }\bibfield  {title}
  {\bibinfo {title} {Electrically tunable low-density superconductivity in a
  monolayer topological insulator},\ }\href
  {https://doi.org/10.1126/science.aar4642} {\bibfield  {journal} {\bibinfo
  {journal} {Science}\ }\textbf {\bibinfo {volume} {362}},\ \bibinfo {pages}
  {926} (\bibinfo {year} {2018})}\BibitemShut {NoStop}%
\bibitem [{\citenamefont {Sajadi}\ \emph {et~al.}(2018)\citenamefont {Sajadi},
  \citenamefont {Palomaki}, \citenamefont {Fei}, \citenamefont {Zhao},
  \citenamefont {Bement}, \citenamefont {Olsen}, \citenamefont {Luescher},
  \citenamefont {Xu}, \citenamefont {Folk},\ and\ \citenamefont
  {Cobden}}]{science.aar4426}%
  \BibitemOpen
  \bibfield  {author} {\bibinfo {author} {\bibfnamefont {E.}~\bibnamefont
  {Sajadi}}, \bibinfo {author} {\bibfnamefont {T.}~\bibnamefont {Palomaki}},
  \bibinfo {author} {\bibfnamefont {Z.}~\bibnamefont {Fei}}, \bibinfo {author}
  {\bibfnamefont {W.}~\bibnamefont {Zhao}}, \bibinfo {author} {\bibfnamefont
  {P.}~\bibnamefont {Bement}}, \bibinfo {author} {\bibfnamefont
  {C.}~\bibnamefont {Olsen}}, \bibinfo {author} {\bibfnamefont
  {S.}~\bibnamefont {Luescher}}, \bibinfo {author} {\bibfnamefont
  {X.}~\bibnamefont {Xu}}, \bibinfo {author} {\bibfnamefont {J.~A.}\
  \bibnamefont {Folk}},\ and\ \bibinfo {author} {\bibfnamefont {D.~H.}\
  \bibnamefont {Cobden}},\ }\bibfield  {title} {\bibinfo {title} {Gate-induced
  superconductivity in a monolayer topological insulator},\ }\href
  {https://doi.org/10.1126/science.aar4426} {\bibfield  {journal} {\bibinfo
  {journal} {Science}\ }\textbf {\bibinfo {volume} {362}},\ \bibinfo {pages}
  {922} (\bibinfo {year} {2018})}\BibitemShut {NoStop}%
\bibitem [{\citenamefont {Zhang}\ \emph {et~al.}(2013)\citenamefont {Zhang},
  \citenamefont {Tanaka}, \citenamefont {Watanabe}, \citenamefont {Zhu},
  \citenamefont {Inumaru},\ and\ \citenamefont {Yamanaka}}]{zhang2013}%
  \BibitemOpen
  \bibfield  {author} {\bibinfo {author} {\bibfnamefont {S.}~\bibnamefont
  {Zhang}}, \bibinfo {author} {\bibfnamefont {M.}~\bibnamefont {Tanaka}},
  \bibinfo {author} {\bibfnamefont {E.}~\bibnamefont {Watanabe}}, \bibinfo
  {author} {\bibfnamefont {H.}~\bibnamefont {Zhu}}, \bibinfo {author}
  {\bibfnamefont {K.}~\bibnamefont {Inumaru}},\ and\ \bibinfo {author}
  {\bibfnamefont {S.}~\bibnamefont {Yamanaka}},\ }\bibfield  {title} {\bibinfo
  {title} {{Superconductivity of alkali metal intercalated TiNBr with
  $\alpha$-type nitride layers}},\ }\href
  {https://doi.org/10.1088/0953-2048/26/12/122001} {\bibfield  {journal}
  {\bibinfo  {journal} {Superconductor Science and Technology}\ }\textbf
  {\bibinfo {volume} {26}},\ \bibinfo {pages} {122001} (\bibinfo {year}
  {2013})}\BibitemShut {NoStop}%
\bibitem [{\citenamefont {Sigrist}\ and\ \citenamefont
  {Ueda}(1991)}]{RevModPhys.63.239}%
  \BibitemOpen
  \bibfield  {author} {\bibinfo {author} {\bibfnamefont {M.}~\bibnamefont
  {Sigrist}}\ and\ \bibinfo {author} {\bibfnamefont {K.}~\bibnamefont {Ueda}},\
  }\bibfield  {title} {\bibinfo {title} {Phenomenological theory of
  unconventional superconductivity},\ }\href
  {https://doi.org/10.1103/RevModPhys.63.239} {\bibfield  {journal} {\bibinfo
  {journal} {Rev. Mod. Phys.}\ }\textbf {\bibinfo {volume} {63}},\ \bibinfo
  {pages} {239} (\bibinfo {year} {1991})}\BibitemShut {NoStop}%
\bibitem [{\citenamefont {Cheng}\ \emph {et~al.}(2010)\citenamefont {Cheng},
  \citenamefont {Sun}, \citenamefont {Galitski},\ and\ \citenamefont
  {Das~Sarma}}]{PhysRevB.81.024504}%
  \BibitemOpen
  \bibfield  {author} {\bibinfo {author} {\bibfnamefont {M.}~\bibnamefont
  {Cheng}}, \bibinfo {author} {\bibfnamefont {K.}~\bibnamefont {Sun}}, \bibinfo
  {author} {\bibfnamefont {V.}~\bibnamefont {Galitski}},\ and\ \bibinfo
  {author} {\bibfnamefont {S.}~\bibnamefont {Das~Sarma}},\ }\bibfield  {title}
  {\bibinfo {title} {Stable topological superconductivity in a family of
  two-dimensional fermion models},\ }\href
  {https://doi.org/10.1103/PhysRevB.81.024504} {\bibfield  {journal} {\bibinfo
  {journal} {Phys. Rev. B}\ }\textbf {\bibinfo {volume} {81}},\ \bibinfo
  {pages} {024504} (\bibinfo {year} {2010})}\BibitemShut {NoStop}%
\bibitem [{\citenamefont {Hesselmann}\ \emph {et~al.}(2018)\citenamefont
  {Hesselmann}, \citenamefont {Scherer}, \citenamefont {Scherer},\ and\
  \citenamefont {Wessel}}]{PhysRevB.98.045142}%
  \BibitemOpen
  \bibfield  {author} {\bibinfo {author} {\bibfnamefont {S.}~\bibnamefont
  {Hesselmann}}, \bibinfo {author} {\bibfnamefont {D.~D.}\ \bibnamefont
  {Scherer}}, \bibinfo {author} {\bibfnamefont {M.~M.}\ \bibnamefont
  {Scherer}},\ and\ \bibinfo {author} {\bibfnamefont {S.}~\bibnamefont
  {Wessel}},\ }\bibfield  {title} {\bibinfo {title} {{Bond-ordered states and
  $f$-wave pairing of spinless fermions on the honeycomb lattice}},\ }\href
  {https://doi.org/10.1103/PhysRevB.98.045142} {\bibfield  {journal} {\bibinfo
  {journal} {Phys. Rev. B}\ }\textbf {\bibinfo {volume} {98}},\ \bibinfo
  {pages} {045142} (\bibinfo {year} {2018})}\BibitemShut {NoStop}%
\bibitem [{\citenamefont {Ma}\ and\ \citenamefont
  {Tong}(2021)}]{PhysRevB.104.155116}%
  \BibitemOpen
  \bibfield  {author} {\bibinfo {author} {\bibfnamefont {K.-H.}\ \bibnamefont
  {Ma}}\ and\ \bibinfo {author} {\bibfnamefont {N.-H.}\ \bibnamefont {Tong}},\
  }\bibfield  {title} {\bibinfo {title} {Interacting spinless fermions on the
  square lattice: Charge order, phase separation, and superconductivity},\
  }\href {https://doi.org/10.1103/PhysRevB.104.155116} {\bibfield  {journal}
  {\bibinfo  {journal} {Phys. Rev. B}\ }\textbf {\bibinfo {volume} {104}},\
  \bibinfo {pages} {155116} (\bibinfo {year} {2021})}\BibitemShut {NoStop}%
\bibitem [{\citenamefont {Gneist}\ \emph {et~al.}(2022)\citenamefont {Gneist},
  \citenamefont {Kiese}, \citenamefont {Henkel}, \citenamefont {Thomale},
  \citenamefont {Classen},\ and\ \citenamefont {Scherer}}]{Gneist2022}%
  \BibitemOpen
  \bibfield  {author} {\bibinfo {author} {\bibfnamefont {N.}~\bibnamefont
  {Gneist}}, \bibinfo {author} {\bibfnamefont {D.}~\bibnamefont {Kiese}},
  \bibinfo {author} {\bibfnamefont {R.}~\bibnamefont {Henkel}}, \bibinfo
  {author} {\bibfnamefont {R.}~\bibnamefont {Thomale}}, \bibinfo {author}
  {\bibfnamefont {L.}~\bibnamefont {Classen}},\ and\ \bibinfo {author}
  {\bibfnamefont {M.~M.}\ \bibnamefont {Scherer}},\ }\bibfield  {title}
  {\bibinfo {title} {Functional renormalization of spinless triangular-lattice
  fermions: N-patch vs. truncated-unity scheme},\ }\href
  {https://doi.org/10.1140/epjb/s10051-022-00395-w} {\bibfield  {journal}
  {\bibinfo  {journal} {Eur. Phys. J. B}\ }\textbf {\bibinfo {volume} {95}},\
  \bibinfo {pages} {157} (\bibinfo {year} {2022})}\BibitemShut {NoStop}%
\bibitem [{\citenamefont {White}(1992)}]{White}%
  \BibitemOpen
  \bibfield  {author} {\bibinfo {author} {\bibfnamefont {S.~R.}\ \bibnamefont
  {White}},\ }\bibfield  {title} {\bibinfo {title} {{Density matrix formulation
  for quantum renormalization groups}},\ }\href
  {https://doi.org/10.1103/PhysRevLett.69.2863} {\bibfield  {journal} {\bibinfo
   {journal} {Phys. Rev. Lett.}\ }\textbf {\bibinfo {volume} {69}},\ \bibinfo
  {pages} {2863} (\bibinfo {year} {1992})}\BibitemShut {NoStop}%
\bibitem [{\citenamefont {McCulloch}(2008)}]{mcculloch2008infinite}%
  \BibitemOpen
  \bibfield  {author} {\bibinfo {author} {\bibfnamefont {I.~P.}\ \bibnamefont
  {McCulloch}},\ }\bibfield  {title} {\bibinfo {title} {{Infinite size density
  matrix renormalization group, revisited}},\ }\href@noop {} {\bibfield
  {journal} {\bibinfo  {journal} {arXiv preprint arXiv:0804.2509}\ } (\bibinfo
  {year} {2008})}\BibitemShut {NoStop}%
\bibitem [{\citenamefont {Metzner}\ \emph {et~al.}(2012)\citenamefont
  {Metzner}, \citenamefont {Salmhofer}, \citenamefont {Honerkamp},
  \citenamefont {Meden},\ and\ \citenamefont
  {Sch\"onhammer}}]{RevModPhys.84.299}%
  \BibitemOpen
  \bibfield  {author} {\bibinfo {author} {\bibfnamefont {W.}~\bibnamefont
  {Metzner}}, \bibinfo {author} {\bibfnamefont {M.}~\bibnamefont {Salmhofer}},
  \bibinfo {author} {\bibfnamefont {C.}~\bibnamefont {Honerkamp}}, \bibinfo
  {author} {\bibfnamefont {V.}~\bibnamefont {Meden}},\ and\ \bibinfo {author}
  {\bibfnamefont {K.}~\bibnamefont {Sch\"onhammer}},\ }\bibfield  {title}
  {\bibinfo {title} {Functional renormalization group approach to correlated
  fermion systems},\ }\href {https://doi.org/10.1103/RevModPhys.84.299}
  {\bibfield  {journal} {\bibinfo  {journal} {Rev. Mod. Phys.}\ }\textbf
  {\bibinfo {volume} {84}},\ \bibinfo {pages} {299} (\bibinfo {year}
  {2012})}\BibitemShut {NoStop}%
\bibitem [{\citenamefont {Read}\ and\ \citenamefont {Green}(2000)}]{RG}%
  \BibitemOpen
  \bibfield  {author} {\bibinfo {author} {\bibfnamefont {N.}~\bibnamefont
  {Read}}\ and\ \bibinfo {author} {\bibfnamefont {D.}~\bibnamefont {Green}},\
  }\bibfield  {title} {\bibinfo {title} {{Paired states of fermions in two
  dimensions with breaking of parity and time-reversal symmetries and the
  fractional quantum Hall effect}},\ }\href
  {https://doi.org/10.1103/PhysRevB.61.10267} {\bibfield  {journal} {\bibinfo
  {journal} {Phys. Rev. B}\ }\textbf {\bibinfo {volume} {61}},\ \bibinfo
  {pages} {10267} (\bibinfo {year} {2000})}\BibitemShut {NoStop}%
\bibitem [{\citenamefont {MacDonald}\ \emph {et~al.}(1988)\citenamefont
  {MacDonald}, \citenamefont {Girvin},\ and\ \citenamefont
  {Yoshioka}}]{MacDonald}%
  \BibitemOpen
  \bibfield  {author} {\bibinfo {author} {\bibfnamefont {A.~H.}\ \bibnamefont
  {MacDonald}}, \bibinfo {author} {\bibfnamefont {S.~M.}\ \bibnamefont
  {Girvin}},\ and\ \bibinfo {author} {\bibfnamefont {D.}~\bibnamefont
  {Yoshioka}},\ }\bibfield  {title} {\bibinfo {title} {{$\frac{t}{U}$ expansion
  for the Hubbard model}},\ }\href {https://doi.org/10.1103/PhysRevB.37.9753}
  {\bibfield  {journal} {\bibinfo  {journal} {Phys. Rev. B}\ }\textbf {\bibinfo
  {volume} {37}},\ \bibinfo {pages} {9753} (\bibinfo {year}
  {1988})}\BibitemShut {NoStop}%
\bibitem [{\citenamefont {Chernyshev}\ \emph {et~al.}(2004)\citenamefont
  {Chernyshev}, \citenamefont {Galanakis}, \citenamefont {Phillips},
  \citenamefont {Rozhkov},\ and\ \citenamefont {Tremblay}}]{Chernyshev}%
  \BibitemOpen
  \bibfield  {author} {\bibinfo {author} {\bibfnamefont {A.~L.}\ \bibnamefont
  {Chernyshev}}, \bibinfo {author} {\bibfnamefont {D.}~\bibnamefont
  {Galanakis}}, \bibinfo {author} {\bibfnamefont {P.}~\bibnamefont {Phillips}},
  \bibinfo {author} {\bibfnamefont {A.~V.}\ \bibnamefont {Rozhkov}},\ and\
  \bibinfo {author} {\bibfnamefont {A.-M.~S.}\ \bibnamefont {Tremblay}},\
  }\bibfield  {title} {\bibinfo {title} {{Higher order corrections to effective
  low-energy theories for strongly correlated electron systems}},\ }\href
  {https://doi.org/10.1103/PhysRevB.70.235111} {\bibfield  {journal} {\bibinfo
  {journal} {Phys. Rev. B}\ }\textbf {\bibinfo {volume} {70}},\ \bibinfo
  {pages} {235111} (\bibinfo {year} {2004})}\BibitemShut {NoStop}%
\bibitem [{sup()}]{suppm}%
  \BibitemOpen
  \href@noop {} {}\bibinfo {note} {See Supplemental Material at
  XXX}\BibitemShut {NoStop}%
\bibitem [{\citenamefont {Castro~Neto}\ \emph {et~al.}(2009)\citenamefont
  {Castro~Neto}, \citenamefont {Guinea}, \citenamefont {Peres}, \citenamefont
  {Novoselov},\ and\ \citenamefont {Geim}}]{Castro_Neto_2009}%
  \BibitemOpen
  \bibfield  {author} {\bibinfo {author} {\bibfnamefont {A.~H.}\ \bibnamefont
  {Castro~Neto}}, \bibinfo {author} {\bibfnamefont {F.}~\bibnamefont {Guinea}},
  \bibinfo {author} {\bibfnamefont {N.~M.~R.}\ \bibnamefont {Peres}}, \bibinfo
  {author} {\bibfnamefont {K.~S.}\ \bibnamefont {Novoselov}},\ and\ \bibinfo
  {author} {\bibfnamefont {A.~K.}\ \bibnamefont {Geim}},\ }\bibfield  {title}
  {\bibinfo {title} {The electronic properties of graphene},\ }\href
  {https://doi.org/10.1103/RevModPhys.81.109} {\bibfield  {journal} {\bibinfo
  {journal} {Rev. Mod. Phys.}\ }\textbf {\bibinfo {volume} {81}},\ \bibinfo
  {pages} {109} (\bibinfo {year} {2009})}\BibitemShut {NoStop}%
\bibitem [{\citenamefont {Dupuis}\ \emph {et~al.}(2021)\citenamefont {Dupuis},
  \citenamefont {Canet}, \citenamefont {Eichhorn}, \citenamefont {Metzner},
  \citenamefont {Pawlowski}, \citenamefont {Tissier},\ and\ \citenamefont
  {Wschebor}}]{Dupuis_2021}%
  \BibitemOpen
  \bibfield  {author} {\bibinfo {author} {\bibfnamefont {N.}~\bibnamefont
  {Dupuis}}, \bibinfo {author} {\bibfnamefont {L.}~\bibnamefont {Canet}},
  \bibinfo {author} {\bibfnamefont {A.}~\bibnamefont {Eichhorn}}, \bibinfo
  {author} {\bibfnamefont {W.}~\bibnamefont {Metzner}}, \bibinfo {author}
  {\bibfnamefont {J.}~\bibnamefont {Pawlowski}}, \bibinfo {author}
  {\bibfnamefont {M.}~\bibnamefont {Tissier}},\ and\ \bibinfo {author}
  {\bibfnamefont {N.}~\bibnamefont {Wschebor}},\ }\bibfield  {title} {\bibinfo
  {title} {The nonperturbative functional renormalization group and its
  applications},\ }\href {https://doi.org/10.1016/j.physrep.2021.01.001}
  {\bibfield  {journal} {\bibinfo  {journal} {Physics Reports}\ }\textbf
  {\bibinfo {volume} {910}},\ \bibinfo {pages} {1} (\bibinfo {year}
  {2021})}\BibitemShut {NoStop}%
\bibitem [{\citenamefont {Markhof}\ \emph {et~al.}(2018)\citenamefont
  {Markhof}, \citenamefont {Sbierski}, \citenamefont {Meden},\ and\
  \citenamefont {Karrasch}}]{Markhof_2018}%
  \BibitemOpen
  \bibfield  {author} {\bibinfo {author} {\bibfnamefont {L.}~\bibnamefont
  {Markhof}}, \bibinfo {author} {\bibfnamefont {B.}~\bibnamefont {Sbierski}},
  \bibinfo {author} {\bibfnamefont {V.}~\bibnamefont {Meden}},\ and\ \bibinfo
  {author} {\bibfnamefont {C.}~\bibnamefont {Karrasch}},\ }\bibfield  {title}
  {\bibinfo {title} {Detecting phases in one-dimensional many-fermion systems
  with the functional renormalization group},\ }\href
  {https://doi.org/10.1103/PhysRevB.97.235126} {\bibfield  {journal} {\bibinfo
  {journal} {Phys. Rev. B}\ }\textbf {\bibinfo {volume} {97}},\ \bibinfo
  {pages} {235126} (\bibinfo {year} {2018})}\BibitemShut {NoStop}%
\bibitem [{\citenamefont {{Profe, Jonas B.}}\ and\ \citenamefont {{Kennes,
  Dante M.}}(2022)}]{Profe_2022}%
  \BibitemOpen
  \bibfield  {author} {\bibinfo {author} {\bibnamefont {{Profe, Jonas B.}}}\
  and\ \bibinfo {author} {\bibnamefont {{Kennes, Dante M.}}},\ }\bibfield
  {title} {\bibinfo {title} {{TU}{$^2$}{FRG}: a scalable approach for truncated
  unity functional renormalization group in generic fermionic models},\ }\href
  {https://doi.org/10.1140/epjb/s10051-022-00316-x} {\bibfield  {journal}
  {\bibinfo  {journal} {Eur. Phys. J. B}\ }\textbf {\bibinfo {volume} {95}},\
  \bibinfo {pages} {60} (\bibinfo {year} {2022})}\BibitemShut {NoStop}%
\bibitem [{\citenamefont {Beyer}\ \emph {et~al.}(2022)\citenamefont {Beyer},
  \citenamefont {Profe},\ and\ \citenamefont {Klebl}}]{Beyer_2022}%
  \BibitemOpen
  \bibfield  {author} {\bibinfo {author} {\bibfnamefont {J.}~\bibnamefont
  {Beyer}}, \bibinfo {author} {\bibfnamefont {J.~B.}\ \bibnamefont {Profe}},\
  and\ \bibinfo {author} {\bibfnamefont {L.}~\bibnamefont {Klebl}},\ }\bibfield
   {title} {\bibinfo {title} {Reference results for the momentum space
  functional renormalization group},\ }\href
  {https://doi.org/10.1140/epjb/s10051-022-00323-y} {\bibfield  {journal}
  {\bibinfo  {journal} {Eur. Phys. J. B}\ }\textbf {\bibinfo {volume} {95}},\
  \bibinfo {pages} {65} (\bibinfo {year} {2022})}\BibitemShut {NoStop}%
\bibitem [{\citenamefont {Sinnott}\ and\ \citenamefont {Andrews}(2001)}]{CNT}%
  \BibitemOpen
  \bibfield  {author} {\bibinfo {author} {\bibfnamefont {S.~B.}\ \bibnamefont
  {Sinnott}}\ and\ \bibinfo {author} {\bibfnamefont {R.}~\bibnamefont
  {Andrews}},\ }\bibfield  {title} {\bibinfo {title} {Carbon nanotubes:
  Synthesis, properties, and applications},\ }\href
  {https://doi.org/10.1080/20014091104189} {\bibfield  {journal} {\bibinfo
  {journal} {Crit. Rev. Solid State Mater. Sci.}\ }\textbf {\bibinfo {volume}
  {26}},\ \bibinfo {pages} {145} (\bibinfo {year} {2001})}\BibitemShut
  {NoStop}%
\bibitem [{\citenamefont {Peng}\ \emph {et~al.}(2022)\citenamefont {Peng},
  \citenamefont {Wang}, \citenamefont {Wen}, \citenamefont {Lee}, \citenamefont
  {Devereaux},\ and\ \citenamefont {Jiang}}]{peng2022enhanced}%
  \BibitemOpen
  \bibfield  {author} {\bibinfo {author} {\bibfnamefont {C.}~\bibnamefont
  {Peng}}, \bibinfo {author} {\bibfnamefont {Y.}~\bibnamefont {Wang}}, \bibinfo
  {author} {\bibfnamefont {J.}~\bibnamefont {Wen}}, \bibinfo {author}
  {\bibfnamefont {Y.}~\bibnamefont {Lee}}, \bibinfo {author} {\bibfnamefont
  {T.}~\bibnamefont {Devereaux}},\ and\ \bibinfo {author} {\bibfnamefont
  {H.-C.}\ \bibnamefont {Jiang}},\ }\bibfield  {title} {\bibinfo {title}
  {{Enhanced superconductivity by near-neighbor attraction in the doped Hubbard
  model}},\ }\href@noop {} {\bibfield  {journal} {\bibinfo  {journal} {arXiv
  preprint arXiv:2206.03486}\ } (\bibinfo {year} {2022})}\BibitemShut {NoStop}%
\bibitem [{\citenamefont {Zauner}\ \emph {et~al.}(2015)\citenamefont {Zauner},
  \citenamefont {Draxler}, \citenamefont {Vanderstraeten}, \citenamefont
  {Degroote}, \citenamefont {Haegeman}, \citenamefont {Rams}, \citenamefont
  {Stojevic}, \citenamefont {Schuch},\ and\ \citenamefont
  {Verstraete}}]{Zauner_2015}%
  \BibitemOpen
  \bibfield  {author} {\bibinfo {author} {\bibfnamefont {V.}~\bibnamefont
  {Zauner}}, \bibinfo {author} {\bibfnamefont {D.}~\bibnamefont {Draxler}},
  \bibinfo {author} {\bibfnamefont {L.}~\bibnamefont {Vanderstraeten}},
  \bibinfo {author} {\bibfnamefont {M.}~\bibnamefont {Degroote}}, \bibinfo
  {author} {\bibfnamefont {J.}~\bibnamefont {Haegeman}}, \bibinfo {author}
  {\bibfnamefont {M.~M.}\ \bibnamefont {Rams}}, \bibinfo {author}
  {\bibfnamefont {V.}~\bibnamefont {Stojevic}}, \bibinfo {author}
  {\bibfnamefont {N.}~\bibnamefont {Schuch}},\ and\ \bibinfo {author}
  {\bibfnamefont {F.}~\bibnamefont {Verstraete}},\ }\bibfield  {title}
  {\bibinfo {title} {Transfer matrices and excitations with matrix product
  states},\ }\href {https://doi.org/10.1088/1367-2630/17/5/053002} {\bibfield
  {journal} {\bibinfo  {journal} {New Journal of Physics}\ }\textbf {\bibinfo
  {volume} {17}},\ \bibinfo {pages} {053002} (\bibinfo {year}
  {2015})}\BibitemShut {NoStop}%
\bibitem [{\citenamefont {He}\ \emph {et~al.}(2021)\citenamefont {He},
  \citenamefont {Pekker},\ and\ \citenamefont {Mong}}]{PhysRevB.104.195126}%
  \BibitemOpen
  \bibfield  {author} {\bibinfo {author} {\bibfnamefont {Y.}~\bibnamefont
  {He}}, \bibinfo {author} {\bibfnamefont {D.}~\bibnamefont {Pekker}},\ and\
  \bibinfo {author} {\bibfnamefont {R.~S.~K.}\ \bibnamefont {Mong}},\
  }\bibfield  {title} {\bibinfo {title} {{One-dimensional repulsive Hubbard
  model with mass imbalance: Orders and filling anomaly}},\ }\href
  {https://doi.org/10.1103/PhysRevB.104.195126} {\bibfield  {journal} {\bibinfo
   {journal} {Phys. Rev. B}\ }\textbf {\bibinfo {volume} {104}},\ \bibinfo
  {pages} {195126} (\bibinfo {year} {2021})}\BibitemShut {NoStop}%
\bibitem [{\citenamefont {Pollmann}\ \emph {et~al.}(2009)\citenamefont
  {Pollmann}, \citenamefont {Mukerjee}, \citenamefont {Turner},\ and\
  \citenamefont {Moore}}]{pollmann2009theory}%
  \BibitemOpen
  \bibfield  {author} {\bibinfo {author} {\bibfnamefont {F.}~\bibnamefont
  {Pollmann}}, \bibinfo {author} {\bibfnamefont {S.}~\bibnamefont {Mukerjee}},
  \bibinfo {author} {\bibfnamefont {A.~M.}\ \bibnamefont {Turner}},\ and\
  \bibinfo {author} {\bibfnamefont {J.~E.}\ \bibnamefont {Moore}},\ }\bibfield
  {title} {\bibinfo {title} {{Theory of Finite-Entanglement Scaling at
  One-Dimensional Quantum Critical Points}},\ }\href
  {https://doi.org/10.1103/PhysRevLett.102.255701} {\bibfield  {journal}
  {\bibinfo  {journal} {Phys. Rev. Lett.}\ }\textbf {\bibinfo {volume} {102}},\
  \bibinfo {pages} {255701} (\bibinfo {year} {2009})}\BibitemShut {NoStop}%
\bibitem [{\citenamefont {Fukui}\ \emph {et~al.}(2005)\citenamefont {Fukui},
  \citenamefont {Hatsugai},\ and\ \citenamefont {Suzuki}}]{Fukui_2005}%
  \BibitemOpen
  \bibfield  {author} {\bibinfo {author} {\bibfnamefont {T.}~\bibnamefont
  {Fukui}}, \bibinfo {author} {\bibfnamefont {Y.}~\bibnamefont {Hatsugai}},\
  and\ \bibinfo {author} {\bibfnamefont {H.}~\bibnamefont {Suzuki}},\
  }\bibfield  {title} {\bibinfo {title} {{Chern Numbers in Discretized
  Brillouin Zone: Efficient Method of Computing (Spin) Hall Conductances}},\
  }\href {https://doi.org/10.1143/jpsj.74.1674} {\bibfield  {journal} {\bibinfo
   {journal} {J. Phys. Soc. Jpn.}\ }\textbf {\bibinfo {volume} {74}},\ \bibinfo
  {pages} {1674} (\bibinfo {year} {2005})}\BibitemShut {NoStop}%
\bibitem [{\citenamefont {Scalapino}\ \emph {et~al.}(1987)\citenamefont
  {Scalapino}, \citenamefont {Loh},\ and\ \citenamefont
  {Hirsch}}]{PhysRevB.35.6694}%
  \BibitemOpen
  \bibfield  {author} {\bibinfo {author} {\bibfnamefont {D.~J.}\ \bibnamefont
  {Scalapino}}, \bibinfo {author} {\bibfnamefont {E.}~\bibnamefont {Loh}},\
  and\ \bibinfo {author} {\bibfnamefont {J.~E.}\ \bibnamefont {Hirsch}},\
  }\bibfield  {title} {\bibinfo {title} {Fermi-surface instabilities and
  superconducting $d$-wave pairing},\ }\href
  {https://doi.org/10.1103/PhysRevB.35.6694} {\bibfield  {journal} {\bibinfo
  {journal} {Phys. Rev. B}\ }\textbf {\bibinfo {volume} {35}},\ \bibinfo
  {pages} {6694} (\bibinfo {year} {1987})}\BibitemShut {NoStop}%
\bibitem [{\citenamefont {Kane}\ \emph {et~al.}(2017)\citenamefont {Kane},
  \citenamefont {Stern},\ and\ \citenamefont {Halperin}}]{KSH}%
  \BibitemOpen
  \bibfield  {author} {\bibinfo {author} {\bibfnamefont {C.~L.}\ \bibnamefont
  {Kane}}, \bibinfo {author} {\bibfnamefont {A.}~\bibnamefont {Stern}},\ and\
  \bibinfo {author} {\bibfnamefont {B.~I.}\ \bibnamefont {Halperin}},\
  }\bibfield  {title} {\bibinfo {title} {{Pairing in Luttinger Liquids and
  Quantum Hall States}},\ }\href {https://doi.org/10.1103/PhysRevX.7.031009}
  {\bibfield  {journal} {\bibinfo  {journal} {Phys. Rev. X}\ }\textbf {\bibinfo
  {volume} {7}},\ \bibinfo {pages} {031009} (\bibinfo {year}
  {2017})}\BibitemShut {NoStop}%
\bibitem [{\citenamefont {He}\ \emph {et~al.}(2019)\citenamefont {He},
  \citenamefont {Tian}, \citenamefont {Pekker},\ and\ \citenamefont
  {Mong}}]{emtll}%
  \BibitemOpen
  \bibfield  {author} {\bibinfo {author} {\bibfnamefont {Y.}~\bibnamefont
  {He}}, \bibinfo {author} {\bibfnamefont {B.}~\bibnamefont {Tian}}, \bibinfo
  {author} {\bibfnamefont {D.}~\bibnamefont {Pekker}},\ and\ \bibinfo {author}
  {\bibfnamefont {R.~S.~K.}\ \bibnamefont {Mong}},\ }\bibfield  {title}
  {\bibinfo {title} {{Emergent mode and bound states in single-component
  one-dimensional lattice fermionic systems}},\ }\href
  {https://doi.org/10.1103/PhysRevB.100.201101} {\bibfield  {journal} {\bibinfo
   {journal} {Phys. Rev. B}\ }\textbf {\bibinfo {volume} {100}},\ \bibinfo
  {pages} {201101} (\bibinfo {year} {2019})}\BibitemShut {NoStop}%
\bibitem [{\citenamefont {Gotta}\ \emph {et~al.}(2021)\citenamefont {Gotta},
  \citenamefont {Mazza}, \citenamefont {Simon},\ and\ \citenamefont
  {Roux}}]{PhysRevResearch.3.013114}%
  \BibitemOpen
  \bibfield  {author} {\bibinfo {author} {\bibfnamefont {L.}~\bibnamefont
  {Gotta}}, \bibinfo {author} {\bibfnamefont {L.}~\bibnamefont {Mazza}},
  \bibinfo {author} {\bibfnamefont {P.}~\bibnamefont {Simon}},\ and\ \bibinfo
  {author} {\bibfnamefont {G.}~\bibnamefont {Roux}},\ }\bibfield  {title}
  {\bibinfo {title} {{Pairing in spinless fermions and spin chains with
  next-nearest neighbor interactions}},\ }\href
  {https://doi.org/10.1103/PhysRevResearch.3.013114} {\bibfield  {journal}
  {\bibinfo  {journal} {Phys. Rev. Research}\ }\textbf {\bibinfo {volume}
  {3}},\ \bibinfo {pages} {013114} (\bibinfo {year} {2021})}\BibitemShut
  {NoStop}%
\bibitem [{\citenamefont {{Long-Yun Xiao and Shun-Li Yu and Wei Wang and
  Zi-Jian Yao and Jian-Xin Li}}(2016)}]{Xiao_2016}%
  \BibitemOpen
  \bibfield  {author} {\bibinfo {author} {\bibnamefont {{Long-Yun Xiao and
  Shun-Li Yu and Wei Wang and Zi-Jian Yao and Jian-Xin Li}}},\ }\bibfield
  {title} {\bibinfo {title} {Possible singlet and triplet superconductivity on
  honeycomb lattice},\ }\href {https://doi.org/10.1209/0295-5075/115/27008}
  {\bibfield  {journal} {\bibinfo  {journal} {Europhys. Lett.}\ }\textbf
  {\bibinfo {volume} {115}},\ \bibinfo {pages} {27008} (\bibinfo {year}
  {2016})}\BibitemShut {NoStop}%
\bibitem [{\citenamefont {Messer}\ \emph {et~al.}(2015)\citenamefont {Messer},
  \citenamefont {Desbuquois}, \citenamefont {Uehlinger}, \citenamefont {Jotzu},
  \citenamefont {Huber}, \citenamefont {Greif},\ and\ \citenamefont
  {Esslinger}}]{PhysRevLett.115.115303}%
  \BibitemOpen
  \bibfield  {author} {\bibinfo {author} {\bibfnamefont {M.}~\bibnamefont
  {Messer}}, \bibinfo {author} {\bibfnamefont {R.}~\bibnamefont {Desbuquois}},
  \bibinfo {author} {\bibfnamefont {T.}~\bibnamefont {Uehlinger}}, \bibinfo
  {author} {\bibfnamefont {G.}~\bibnamefont {Jotzu}}, \bibinfo {author}
  {\bibfnamefont {S.}~\bibnamefont {Huber}}, \bibinfo {author} {\bibfnamefont
  {D.}~\bibnamefont {Greif}},\ and\ \bibinfo {author} {\bibfnamefont
  {T.}~\bibnamefont {Esslinger}},\ }\bibfield  {title} {\bibinfo {title}
  {{Exploring Competing Density Order in the Ionic Hubbard Model with Ultracold
  Fermions}},\ }\href {https://doi.org/10.1103/PhysRevLett.115.115303}
  {\bibfield  {journal} {\bibinfo  {journal} {Phys. Rev. Lett.}\ }\textbf
  {\bibinfo {volume} {115}},\ \bibinfo {pages} {115303} (\bibinfo {year}
  {2015})}\BibitemShut {NoStop}%
\bibitem [{\citenamefont {Bouadim}\ \emph {et~al.}(2007)\citenamefont
  {Bouadim}, \citenamefont {Paris}, \citenamefont {H\'ebert}, \citenamefont
  {Batrouni},\ and\ \citenamefont {Scalettar}}]{PhysRevB.76.085112}%
  \BibitemOpen
  \bibfield  {author} {\bibinfo {author} {\bibfnamefont {K.}~\bibnamefont
  {Bouadim}}, \bibinfo {author} {\bibfnamefont {N.}~\bibnamefont {Paris}},
  \bibinfo {author} {\bibfnamefont {F.}~\bibnamefont {H\'ebert}}, \bibinfo
  {author} {\bibfnamefont {G.~G.}\ \bibnamefont {Batrouni}},\ and\ \bibinfo
  {author} {\bibfnamefont {R.~T.}\ \bibnamefont {Scalettar}},\ }\bibfield
  {title} {\bibinfo {title} {{Metallic phase in the two-dimensional ionic
  Hubbard model}},\ }\href {https://doi.org/10.1103/PhysRevB.76.085112}
  {\bibfield  {journal} {\bibinfo  {journal} {Phys. Rev. B}\ }\textbf {\bibinfo
  {volume} {76}},\ \bibinfo {pages} {085112} (\bibinfo {year}
  {2007})}\BibitemShut {NoStop}%
\bibitem [{\citenamefont {Watanabe}\ and\ \citenamefont
  {Ishihara}(2013)}]{JPSJ}%
  \BibitemOpen
  \bibfield  {author} {\bibinfo {author} {\bibfnamefont {T.}~\bibnamefont
  {Watanabe}}\ and\ \bibinfo {author} {\bibfnamefont {S.}~\bibnamefont
  {Ishihara}},\ }\bibfield  {title} {\bibinfo {title} {{Band and Mott
  Insulators and Superconductivity in Honeycomb-Lattice Ionic-Hubbard Model}},\
  }\href {https://doi.org/10.7566/JPSJ.82.034704} {\bibfield  {journal}
  {\bibinfo  {journal} {J. Phys. Soc. Jpn.}\ }\textbf {\bibinfo {volume}
  {82}},\ \bibinfo {pages} {034704} (\bibinfo {year} {2013})}\BibitemShut
  {NoStop}%
\bibitem [{\citenamefont {Chattopadhyay}\ \emph {et~al.}(2019)\citenamefont
  {Chattopadhyay}, \citenamefont {Bag}, \citenamefont {Krishnamurthy},\ and\
  \citenamefont {Garg}}]{PhysRevB.99.155127}%
  \BibitemOpen
  \bibfield  {author} {\bibinfo {author} {\bibfnamefont {A.}~\bibnamefont
  {Chattopadhyay}}, \bibinfo {author} {\bibfnamefont {S.}~\bibnamefont {Bag}},
  \bibinfo {author} {\bibfnamefont {H.~R.}\ \bibnamefont {Krishnamurthy}},\
  and\ \bibinfo {author} {\bibfnamefont {A.}~\bibnamefont {Garg}},\ }\bibfield
  {title} {\bibinfo {title} {{Phase diagram of the half-filled ionic Hubbard
  model in the limit of strong correlations}},\ }\href
  {https://doi.org/10.1103/PhysRevB.99.155127} {\bibfield  {journal} {\bibinfo
  {journal} {Phys. Rev. B}\ }\textbf {\bibinfo {volume} {99}},\ \bibinfo
  {pages} {155127} (\bibinfo {year} {2019})}\BibitemShut {NoStop}%
\bibitem [{\citenamefont {Chattopadhyay}\ \emph {et~al.}(2021)\citenamefont
  {Chattopadhyay}, \citenamefont {Krishnamurthy},\ and\ \citenamefont
  {Garg}}]{10.21468/SciPostPhysCore.4.2.009}%
  \BibitemOpen
  \bibfield  {author} {\bibinfo {author} {\bibfnamefont {A.}~\bibnamefont
  {Chattopadhyay}}, \bibinfo {author} {\bibfnamefont {H.~R.}\ \bibnamefont
  {Krishnamurthy}},\ and\ \bibinfo {author} {\bibfnamefont {A.}~\bibnamefont
  {Garg}},\ }\bibfield  {title} {\bibinfo {title} {{Unconventional
  superconductivity in a strongly correlated band-insulator without doping}},\
  }\href {https://doi.org/10.21468/SciPostPhysCore.4.2.009} {\bibfield
  {journal} {\bibinfo  {journal} {SciPost Phys. Core}\ }\textbf {\bibinfo
  {volume} {4}},\ \bibinfo {pages} {9} (\bibinfo {year} {2021})}\BibitemShut
  {NoStop}%
\bibitem [{\citenamefont {Weinberg}\ and\ \citenamefont
  {Bukov}(2019)}]{10.21468/SciPostPhys.7.2.020}%
  \BibitemOpen
  \bibfield  {author} {\bibinfo {author} {\bibfnamefont {P.}~\bibnamefont
  {Weinberg}}\ and\ \bibinfo {author} {\bibfnamefont {M.}~\bibnamefont
  {Bukov}},\ }\bibfield  {title} {\bibinfo {title} {{QuSpin: a Python Package
  for Dynamics and Exact Diagonalisation of Quantum Many Body Systems. Part II:
  bosons, fermions and higher spins}},\ }\href
  {https://doi.org/10.21468/SciPostPhys.7.2.020} {\bibfield  {journal}
  {\bibinfo  {journal} {SciPost Phys.}\ }\textbf {\bibinfo {volume} {7}},\
  \bibinfo {pages} {20} (\bibinfo {year} {2019})}\BibitemShut {NoStop}%
\bibitem [{\citenamefont {Hauschild}\ and\ \citenamefont
  {Pollmann}(2018)}]{10.21468/SciPostPhysLectNotes.5}%
  \BibitemOpen
  \bibfield  {author} {\bibinfo {author} {\bibfnamefont {J.}~\bibnamefont
  {Hauschild}}\ and\ \bibinfo {author} {\bibfnamefont {F.}~\bibnamefont
  {Pollmann}},\ }\bibfield  {title} {\bibinfo {title} {{Efficient numerical
  simulations with Tensor Networks: Tensor Network Python (TeNPy)}},\ }\href
  {https://doi.org/10.21468/SciPostPhysLectNotes.5} {\bibfield  {journal}
  {\bibinfo  {journal} {SciPost Phys. Lect. Notes}\ ,\ \bibinfo {pages} {5}}
  (\bibinfo {year} {2018})}\BibitemShut {NoStop}%
\bibitem [{\citenamefont {{Th{\"o}rnig, Philipp}}(2021)}]{JUWELS}%
  \BibitemOpen
  \bibfield  {author} {\bibinfo {author} {\bibnamefont {{Th{\"o}rnig,
  Philipp}}},\ }\bibfield  {title} {\bibinfo {title} {{JURECA: Data Centric and
  Booster Modules implementing the Modular Supercomputing Architecture at
  J{\"u}lich Supercomputing Centre}},\ }\href@noop {} {\bibfield  {journal}
  {\bibinfo  {journal} {Journal of large-scale research facilities JLSRF}\
  }\textbf {\bibinfo {volume} {7}},\ \bibinfo {pages} {A182} (\bibinfo {year}
  {2021})}\BibitemShut {NoStop}%
\end{thebibliography}%

\renewcommand{\theequation}{S\arabic{equation}}
\setcounter{equation}{0}
\renewcommand{\thefigure}{S\arabic{figure}}
\setcounter{figure}{0}
\renewcommand{\thetable}{S\arabic{table}}
\setcounter{table}{0}
\begin{widetext}

\section{Supplemental Material}

\subsection{Schrieffer-Wolff transformation}
The ground state at filling $n=1$ is given by all $A$ sites occupied. We use the Schrieffer-Wolff transformation to find the low-energy physics when extra electrons are brought in by doping. The Hamiltonian is comprised of a kinetic part $H_k$ which creates high-energy excitation and an interaction part $H_U$ which controls the zero-doping state. The Schrieffer-Wolff transformation is to eliminate the high-energy kinetic Hamiltonian order by order through a unitary transformation $\exp(iS)$
\begin{equation}
    H'=e^{iS}He^{-iS}=H+[iS,H]+\frac{1}{2}[iS,[iS,H]]+\dots.
\end{equation}
The transformation operator is expanded in the order of $t/V$ or $t/U$, $S=S^{1}+S^{(2)}+\dots$. For this purpose, we decompose the kinetic Hamiltonian into the terms preserving the interaction $H^{(0)}_{00}$ and those creating higher energy excitations $\tilde H_k$. The aim of $S^{(1)}$ is to eliminate those components $\tilde H_k$
\begin{equation}
    [iS^{(1)},H_U]=-\tilde H_k,\quad H_k=H^{(0)}_{00}+\tilde H_k \textrm{ and }[\tilde H_k,H_U]\ne 0, [H^{(0)}_{00},H_U]=0\ .
\end{equation}
The SW transformed Hamiltonian to the order of $t^2/U$ is given by the sum $[iS^{(1)},H_k]+[iS^{(1)},[iS^{(1)},H_U]]/2=[iS^{(1)},\tilde H_k]/2+[iS^{(1)},H^{(0)}_{00}]$. The effective theory is obtained by projecting this expression to the state with $A$ occupied.

To obtain $S^{(1)}$, it is more convenient to decompose $H_k$ into different ladder operators of the onsite potential and the nearest-neighbor repulsion. We do this first for the hopping process from an $A$ site to a nearby $B$ site. It increases the onsite potential by $D$. The nearest-neighbor repulsion brought by this process depends on the number of neighbors of $A$ and $B$, $\delta E=(n_B-n_A)V$. With these observations, the kinetic Hamiltonian is decomposed as
\begin{align}
    H_k=&\left(H^{(0)}_{00}+H_A\right)+\left(\sum_{m,n}H^+_{m,n}+H^-_{m,n}\right)\nonumber\\
    =&\left(t'\sum_{ij}c^\dagger_{i,B}c_{j,B}+t'\sum_{ij}c^\dagger_{i,A}c_{j,A}\right)+ \left( t\sum_{j,r,m,n} P^B_{m,j+r}c^\dagger _{j+r,B}c_{j,A}P^A_{n,j}+t\sum_{j,r,m,n} P^A_{m,j-r} c^\dagger_{j-r,A} c_{j,B} P^B_{n,j}\right),\label{eq_dechk}
\end{align}
where the next-nearest-neighbor hopping includes terms preserving $H_U$ and those connecting states with different distributions of $A$ particles. The operators $P^{A/B}_{m,j}$ are projection into the state where the particle $A/B$ at unit cell $j$ has $m$ neighbors occupied. They can be written in terms of the sum over the product operators $\prod_r n_r\prod_{r'} (1-n_{r'})$, where $r,r'$ are the occupied/unoccupied neighbors. Notice that $m,n$ cannot be larger than $N-1$, where $N$ is the number of nearest neighbors in this lattice, as the hopping operators in the middle of Eq.~\eqref{eq_dechk} always eliminate one neighbor. One can verify the following commutation relations
\begin{equation}
    [H_U,H^\pm_{m,n}]=\left[\pm D+(m-n)V\right] H^\pm_{m,n}.
\end{equation}
With these relations, we can define the leading order transformation for the $AB$ hopping to be 
\begin{equation}
    iS^{(1)}_{AB}=\sum_{m,n}\frac{H^+_{m,n}}{D+(m-n)V}-\frac{H^-_{m,n}}{D-(m-n)V}.
\end{equation}

Such similar expressions $S^{(1)}_{AA},S^{(1)}_{BB}$ can also be obtained for the $AA,BB$ hopping with different numbers of neighbors occupied. So $S^{(1)}$ is comprised of operators that create excitations of $H_U$. The observation is that those $S^{(1)}_{AA},S^{(1)}_{BB}$ terms annihilate the state with all $A$ sites occupied. So their contribution to the second order expansion $[iS^{(1)},\tilde H_k]/2$ vanishes after the projection. What are left are terms diagonal in $H_U$. The resulting expression $[iS^{(1)},H^{(0)}_{00}]\simeq t't/U$ is off-diagonal in the ground state of $H_U$. It will be further eliminated to the order $1/U^2$ by the second-order SW transformation. So we can neglect it at the order of $1/U$. The Hamiltonian is simplified to a quadratic form of $H^{\pm}_{m,n}$. As the low energy physics is obtained by projecting $H'$ to the state with all $A$ sites occupied, this requires the total excitation should have equal numbers of $+$ and $-$ and the sum of $m-n$ should vanish. As $H^-_{m,n}$ annihilates the low-energy manifold, it ends up with the following equation
\begin{equation}
    H'_{\textrm{eff}}=-\sum_{m',n}\frac{H^-_{m',m'+N-n}H^+_{N,n}}{D+(N-n)V}.
\end{equation}
As the occupation on site $A$ must be conserved, there are two situations in the above summation. When the bond operators in $H^-_{m',m'+N-n}$ and $H^+_{N,n}$ are taking the same one, we obtain density interaction terms. When they differ, we have hopping terms for $B$ fermions. It is more convenient to write out the Hamiltonian for the four neighbors around one $A$ site. Choosing $(i,j,k,l)$ to be the four neighbors of a particle $A$, we have the following terms for the density interaction part
\begin{align}
    H^U_4=&-\frac{4t^2}{D+3V}(1-n_{i})(1-n_{j})(1-n_{k})(1-n_{l}),\\
    H^U_3=&-\frac{3t^2}{D+2V}(1-n_i)(1-n_j)(1-n_k)n_l+\mathrm{PM},\\
    H^U_2=&-\frac{2t^2}{D+V}(1-n_{i})(1-n_{j})n_kn_l+\mathrm{PM},\\
    H^U_1=&-\frac{t^2}{D}(1-n_{i})n_jn_kn_l+\mathrm{PM},
\end{align}
where $\mathrm{PM}$ means \emph{distinct} combinations obtained by permuting $i,j,k,l$. For the hopping processes from $i\to j$, we need that the $A$ particle has neighbor $i$ occupied before the hopping and $j$ occupied after the hopping. These terms are given by
\begin{align}
    H^k_3=&\frac{t^2}{D+2V}c^\dagger_{i,B}c_{j,B}(1-n_k)(1-n_l)+\mathrm{PM},\\
    H^k_2=&\frac{t^2}{D+V}c^\dagger_{i,B}c_{j,B}n_k(1-n_l)+\mathrm{PM},\\
    H^k_1=&\frac{t^2}{D}c^\dagger_{i,B}c_{j,B}n_kn_l+\mathrm{PM}.
\end{align}
Similarly, $\mathrm{PM}$ means \emph{distinct} combinations obtained by permuting $i,j,k,l$

Now we collect the contributions together. We have two-body, three-body and four-body interactions:
\begin{equation}
    H'_{U,\textrm{eff}}=\sum_{ij} U_2n_in_j+\sum_{ijk} U_3n_in_jn_k+\sum_{ijkl}U_4n_in_jn_kn_l.
\end{equation}
The summation of $i,j,k,l$ is defined by counting the different two-, three- and four-combinations of the neighbors around every $A$ site. So the two-body interaction along the diagonal of the square lattice should be counted twice. Their coefficients are given by
\begin{align}
    U_2&=-\frac{2t^2}{D+V}+\frac{6t^2}{D+2V}-\frac{4t^2}{D+3V},\\
    U_3&=-\frac{t^2}{D}+\frac{6t^2}{D+V}-\frac{9t^2}{D+2V}+\frac{4t^2}{D+3V},\\
    U_4&=\frac{4t^2}{D}-\frac{12t^2}{D+V}+\frac{12t^2}{D+2V}-\frac{4t^2}{D+3V}.
\end{align}
The effective hopping Hamiltonian can be assisted by the other two neighbors around each $A$ site
\begin{equation}
     H'_{k,\textrm{eff}}=\sum_{ij} \lambda_0c^\dagger_{i,B}c_{j,B}+\sum_{ijk} \lambda_1 c^\dagger_{i,B}c_{j,B}n_k +\sum_{ijkl}\lambda_2c^\dagger_{i,B}c_{j,B}n_kn_l,
\end{equation}
where the hopping parameters are
\begin{align}
    \lambda_0=&\frac{t^2}{D+2V}\\
    \lambda_1=&\frac{t^2}{D+V}-\frac{t^2}{D+2V}\\
    \lambda_2=&\frac{t^2}{D}-\frac{2t^2}{D+V}+\frac{t^2}{D+2V}.
\end{align}
All the parameters as a function of $V/D$ are plotted in Fig.~\ref{fig:parameters}.

\end{widetext}
\subsection{Continuum theory for low doping}
\begin{figure}
    \centering
    \includegraphics[width=1\columnwidth]{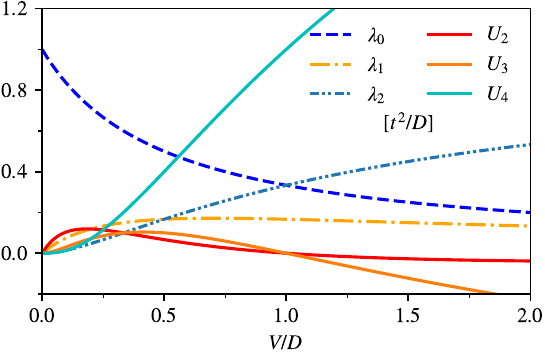}
    \caption{Parameters of the large-D effective theory of the square lattice model.}
    \label{fig:parameters}
\end{figure}
The effective inverse mass tensor at the two valleys are
\begin{equation}
    m^{-1}_\pm=a^2\left(\begin{array}{cc}
       4t'_B \mp 2t_B  & 0\\
       0  & 4t'_B \pm 2t_B,
    \end{array}\right)
\end{equation}
where the subscript $+$ ($-$) denotes the valley located at $(0,\pi)$ ($(\pi,0)$). Recall that the two valleys being  minimum is given by the condition $|t'_B/t_B|>0.5$. We see that the mass tensor is diagonal in the coordinate we choose, and there is mass anisotropy  for each valley.

\begin{figure}
    \centering
    \includegraphics[width=1\columnwidth]{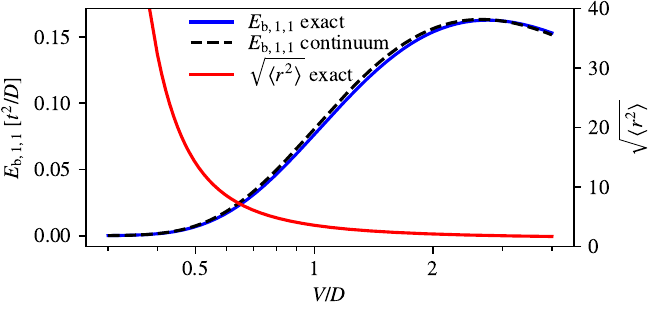}
    \caption{Binding energies and bound states sizes of two fermions for the effective square lattice model with two valleys. The dashed line denotes the continuum effective theory with one fitting parameter given by matching small $V/D$ data.}
    \label{fig:EbR}
\end{figure}

We introduce the center-of-mass coordinates: $\delta \bold{r}= \bold{r}_+- \bold{r}_-, \bold{R}= \bar{m}_{+}\bold{r}_++ \bar{m}_-\bold{r}_-$, where $\bar{m}_{\pm}=m_{\pm}/ \operatorname{Tr}(m_{\pm})$. The (first-quantized) kinetic Hamiltonian of the two particles can be written as $(\nabla^{T}_{\delta \bold{r}}\mu^{-1}\bold{\nabla_{\delta \bold{r}}}+\nabla^{T}_{ \bold{R}} M^{-1}\bold{\nabla_{\bold{R}}})/2$. We find that the relative inverse mass tensor $\mu^{-1}$ is isotropic: $\mu^{-1}=\mu_0^{-1} I$, where $\mu_0^{-1}=(8a^2 t'_B)$ and  $I$ is the $2\times2$ identity matrix. The continuum approximation of the two-particle problem is similar to that of the honeycomb model~\cite{sciadv}. The two-particle binding energies can be calculated in the center-of-mass reference frame~\cite{sciadv}: 
\begin{align}
E_{\mathrm{bp}}=e_{\mathrm{bp}}[e^{2\pi/(\mu_0|g|)}-1]/\mu_0,
\end{align}
where $\mu_0=(8a^2 t'_B)^{-1}$ is the relative mass. The result is independent of the specific value of $t_B$ ($t'$ in the full model) as long as the ground states of single fermion [two fermions] are in the $(0,\pi)$ or $(\pi,0)$ [$(\pi,\pi)$] momentum sector, consistent with the microscopic Hamiltonian being independent of $t_B$ in the two-particle $(\pi,\pi)$ sector.  We fit $e_{\mathrm{bp}}$ using exact data in the small $V/D$ region and plot the continuum effective result together with the exact result in Fig.~\ref{fig:EbR}. Different from the honeycomb model, the binding energies will drop for large $V$ because $|g|$ will decay to zero for large $V$ instead of converging to a constant.

To extract coefficients of interaction in the continuum theory, we write the correlated hopping as
\begin{align}
    \lambda_1 \sum_{i,j,k}c^\dagger_i c^\dagger_jc_j c_k=&\lambda_1 \sum_{\substack{i,j,k,\\\mathbf k,\mathbf k',\mathbf q,\sigma_i}} \psi^\dagger_{\sigma_1,\mathbf  k-\mathbf q} \psi^\dagger_{\sigma_2,\mathbf  k'+\mathbf q}\psi_{\sigma_3,\mathbf  k'} \psi_{\sigma_4,\mathbf k}\nonumber\\
    &\times e^{-i(\mathbf K_{\sigma_1}+\mathbf k-\mathbf q)\cdot \mathbf a_i-i(\mathbf K_{\sigma_2}-\mathbf K_{\sigma_3}+\mathbf q)\cdot \mathbf a_j}\nonumber\\
    &\times e^{i(\mathbf K_{\sigma_4}+\mathbf k)\cdot \mathbf a_k},
\end{align}
where $\mathbf a_i$ is taken from the four vectors connecting $A$ to its nearest $B$ neighbours. The variables $\mathbf k,\mathbf q,\mathbf k'$ are taken to be much smaller than $|\mathbf K^+-\mathbf K^-|$. Similarly, the repulsion term is rewritten as 
\begin{align}
    U_2\sum_{ij}c^\dagger_ic^\dagger_jc_jc_i=&U_2\sum_{\substack{i,j,\mathbf q,\\ \mathbf k,\mathbf k',\sigma_i}}\psi^\dagger_{\sigma_1,\mathbf  k-\mathbf q} \psi^\dagger_{\sigma_2,\mathbf  k'+\mathbf q}\psi_{\sigma_3,\mathbf  k'} \psi_{\sigma_4,\mathbf k}\nonumber\\
    &\times e^{-i(\mathbf K_{\sigma_1}-\mathbf K_{\sigma_4}-\mathbf q)\cdot \mathbf a_i-i(\mathbf K_{\sigma_2}-\mathbf K_{\sigma_3}+\mathbf q)\cdot \mathbf a_j}.
\end{align}
The continuum interactions between different valleys and inside the same valley are given by taking appropriate combinations of $\sigma_i$ and their anti-symmetrized partners.

The result of inter-valley interaction has been given in the main text. Now we consider intra-valley interaction with finite doping. We consider the weak interacting limit and discuss the interaction between modes on the Fermi surface of a valley, for example, the $+$ valley. Here the momenta are defined as the deviation to $(0,\pi)$. In general, we have.
\begin{align}
\sum_{\mathbf k,\mathbf k',\mathbf q} g(\mathbf  k, \mathbf  k', \mathbf  q)\psi^\dagger_{+,\mathbf  k-\mathbf q} \psi^\dagger_{+,\mathbf  k'+\mathbf q}\psi_{+,\mathbf  k'} \psi_{+,\mathbf k}    
\end{align}
We focus on two-particle scattering with net zero deviation to $\mathbf K_{+}$ and low doping (momenta is small enough to perform Taylor expansion).
\begin{align}
\sum_{\mathbf q_1,\mathbf q_2} \tilde{g}(\mathbf  q_1, \mathbf  q_2)\psi^\dagger_{+,\mathbf q_1} \psi^\dagger_{+,-\mathbf q_1}\psi_{+,-\mathbf q_2} \psi_{+,\mathbf q_2}    
\end{align}
Set $\mathbf q_1=\mathbf q_2$ and $\mathbf q_1=-\mathbf q_2$ respectively, we can obtain the density-density interaction, $\propto [2\lambda_1 (q_{1,x}^2-q_{1,y}^2)+ U_2 (q_{1,x}^2+q_{1,y}^2)]n_+(\mathbf q_1)n_+(-\mathbf q_1)$. With the Fermi surface shape close to an eclipse with the long axis along the y direction, and $\lambda_1 \approx U_2$ for $V/D <1$ (Fig.~\ref{fig:parameters}),  such interaction in most momenta is attractive.

\subsection{Few-particle-doping binding energy with finite $D/t$}
\begin{figure}
    \centering
    \includegraphics[width=\columnwidth]{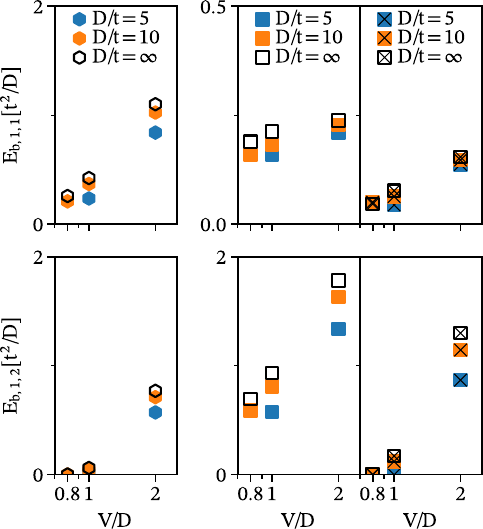}
    \caption{Binding energies. The binding energies of two fermions are shown in the first row from left to right for the honeycomb lattice model  and the square lattice model without $t'$ term and for $t'=\lambda_0$. Correspondingly, the binding energies for a fermion pair and a fermion to form a three-fermion bound state are shown in the second row. The plot scale of the $E_{b,1,1}$ of square lattice models is smaller than others'.}
    \label{fig:bind energies}
\end{figure}

\begin{figure}
    \centering
    \includegraphics[width=\columnwidth]{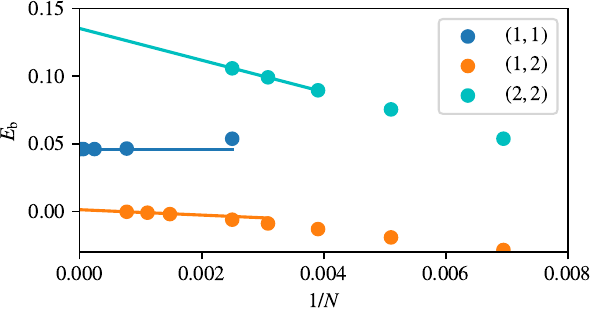}
    \caption{Binding energy extrapolation. We show an example of extrapolation to obtain Fig.~3 in the main text  and Fig.~\ref{fig:bind energies} The data are taken from ED for $V/D=0.8$,  $t'=\lambda_0$ of the square lattice.}
    \label{fig:Ebextraploation}
\end{figure}
Here, we discuss few-particle-doping binding energies of the full model with finite $D/t=10,5$. The binding energies, in the unit of $t^2/D$, are in general smaller for smaller $D/t$ . However, even for $D/t=5$, we find no substantial difference for inferred stable pairing region, compared to the effective model with $D/t=\infty$.

As an alternate of binding energy per particle, we represent the results as binding energies for forming bound states with composites. For two-particle and three-particle bound states, the existence of bound states can be seen from a positive $E_{\mathrm{b},1,1}= 2E_1-E_0-E_2$ and  $E_{\mathrm{b},1,2}= E_1+E_2-E_0-E_3$ respectively. These quantities are plotted in Fig.~\ref{fig:bind energies}. Binding energy per particle can be deduced from them. The existence of three-particle bound states does not mean that three-particle bound states are more favored than pairs for dilute doping. Favored bound states have the largest binding energy per particle.

Some details of the numerical implementation are as follows. For the full (effective) lattice models, we obtain the ground-state energies $E_n$ of finite systems using DMRG (exact) diagonalization. To accurately compute binding energies, we need system sizes larger than the sizes of the bound states.  We estimate the finite-size errors by doing $1/N$ extrapolation for data of the two largest systems we obtain. The extrapolated data in Fig.~\ref{fig:bind energies} have errors smaller than the size of markers. The periodic boundary condition is implemented in the  exact diagonalization, which enables reading the momentum quantum numbers. DMRG is less efficient to deal with periodic conditions along two directions and we thus only implement a periodic boundary condition in one direction while implementing an open boundary condition on the other. In this case, to correctly calculate the bulk binding energy, we find it essential to eliminate the low-energy edge modes. Such modes can be understood by considering the potential and interaction part of the Hamiltonian Eq.~\ref{full model}:$\sum_{\langle i,j \rangle} Vn_in_j+\sum_{i \in B}Dn_i$. As $A$ sublattice is almost fully-filled, if a fermion on B sublattice  is located at the open boundary rather than in the bulk, it feels less repulsion from the fermions on $B$ sublattice. Consequently, these configurations have lower energy. We find that introducing additional potential terms $Vn_i$ on the boundary of $B$ sublattice can eliminate low-energy edge modes.

\subsection{Weak-coupling results}

\begin{figure}
    \centering
    \includegraphics[width=\columnwidth]{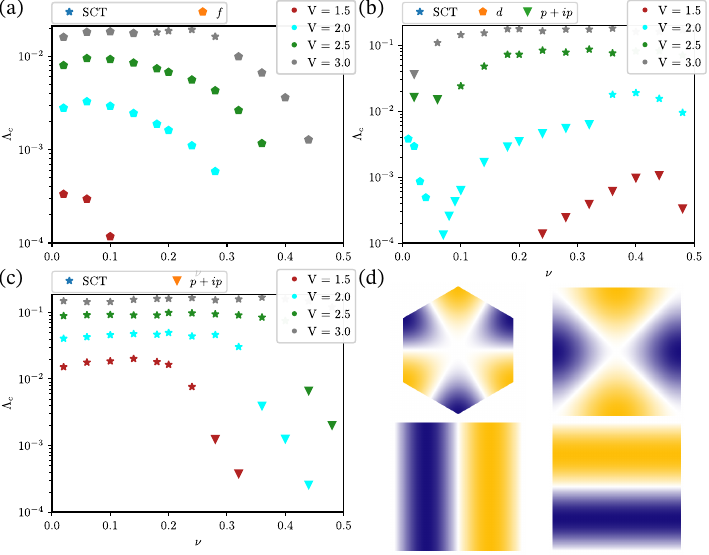}
    \caption{Results of the FRG simulations for the three different setups and visualization of the $f$-wave superconductivity. We abbreviate a flow to strong coupling without divergent susceptibility as SCT, and a divergence of the $f$-wave component of the pairing susceptibility as $f$-SC. The $y$-axis displays the critical energy scale, which is linearly dependent on the critical temperature. The $x$-axis displays the doping. (a) shows the results for a square lattice with $t' = \lambda_0$ and (b) the square lattice with $t'=0$ results. (c) shows the results for the honeycomb lattice. In all simulations, we kept $D = 10$ and varied $V$ in the given range. (d) visualizes the different superconducting order parameter symmetries encountered, as visualized by the eigenvector of the effective two-particle interaction at the orbital at the Fermi level. The upper left shows an $f$-wave on the honeycomb lattice, the upper right shows a $d_{x^2-y^2}$-wave on the square lattice. The lower two plots correspond to the degenerate pair $p_{x}$ and $p_{y}$ with weak admixture of other dependencies.}
    \label{fig:my_label}
\end{figure}

\begin{figure}
    \centering
    \includegraphics[width=0.5\columnwidth]{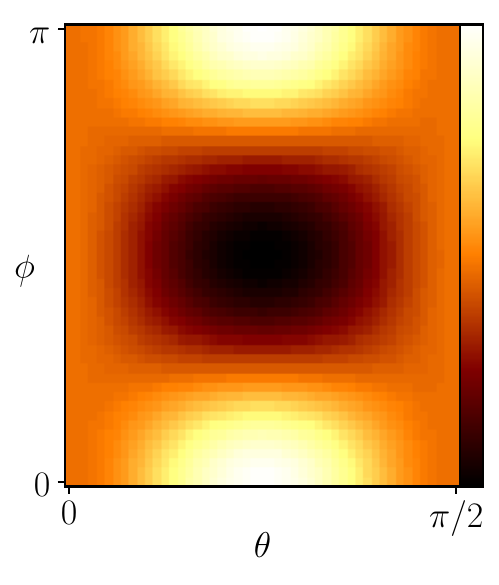}
    \caption{Free energy minimization in the square lattice with $t' = \lambda_0$, using $\cos(\theta)p_{x}+e^{i\phi}\sin(\theta)p_{y}$ as starting values. The combination with the smallest free energy is $\frac{1}{2}\left(p_{x}+ip_{y}\right)$}
    \label{fig:freeenergy}
\end{figure}

In the weak coupling regime, we apply a truncated unity functional renormalization group approach. The FRG flavor we employ is a method to construct unbiased single and two-particle interactions from a set of flow equations. While three- and four-body interactions are not counted as objects themselves, they are partially included via virtual processes in the two-particle interaction.
We use a sharp energy cutoff~\cite{RevModPhys.84.299}, thus the critical energy scale can be interpreted as a critical temperature modulo an unknown scaling factor. We calculate the vertex on a $24\times 24$ momentum mesh for both lattices, with a refinement for the bubble integration mesh of $45 \times 45$. On the square lattice, we include the 25/29 nearest neighbors in the truncated unity per site within the unit cell for the honeycomb/square lattice. We use a Bogacki–Shampine adaptive integrator for the integration of the flow equations, allowing for a maximal absolute error of $10^{-2}$ per integration step. The results of the FRG simulations are visualized in Fig.~\ref{fig:my_label}. To distinguish different phases, we inspect the behavior of the maximal eigenvalues of each channel during the flow in combination with an inspection of the dominant eigenvectors at the end of the flow. These eigenvectors encode symmetries and specific types of instability. In the case of $p_{y}/p_{x}$ we find the two eigenvectors to be exactly degenerate. In the real space representation we define $p_{x/y} = \textrm{sign}(\vec{v}_{x/y}\cdot \vec{d}) \delta_{\vec{v}_{x/y}, \vec{d}}$ with $\vec{v}_{x} = (1,0)$, $\vec{v}_{y} = (0,1)$ and $\vec{d}$ is the vectorial distance between two sites. To distinguish all possible linear combinations $\cos(\theta)p_{x}+e^{i\phi}\sin(\theta)p_{y}$ we perform a single-step mean-field calculation and compare the free energy of each starting configuration, as can be seen in Fig.~\ref{fig:freeenergy}. To calculate the Chern number in the gapped phase, we employ the method described in Ref.~\cite{Fukui_2005}.

\subsection{Details of infinite DMRG calculations}

We obtain approximate ground states of the Hamiltonian Eq.~\ref{full model} defined on infinite cylinders. We do this by optimizing infinite matrix product states via two-site iDMRG algorithm~\cite{mcculloch2008infinite}. We implement the  conservation of particle numbers, thus the phase diagram Fig.~\ref{fig:pd} is constructed in terms of doping densities $\nu$. The accuracy of infinite matrix product states can be improved by increasing its bond dimensions ($\chi$, size of the matrices); With efficient optimization, to reach a given accuracy, the required  computational resource (e.g. $\chi$) is exponentially large in cylinder circumference.  Infinite matrix product states are constructed to be exactly translationally invariant by $M$ lattice unit vector along the axial direction. To implement exact particle number conservation for doping density $\nu=p/q$ (irreducible fraction), $M$ must be integer multiples of $q/L_y$. ($L_y$ is the number of lattice unit vectors around the cylinder.) To be compatible with the possible spontaneous breaking of translational symmetry (e.g., charge-density-wave state), $M$ has to be compatible with the enlarged unit cell.  As mentioned in the main text, we estimate of correlation lengths of single-particle and pair to infer superconductivity. See Ref.~\cite{PhysRevB.104.195126,Zauner_2015} for the definition and extraction methods for one-dimensional correlation lengths. Here, we define the correlation length from the correlation along the axial (infinite) direction. We denote them estimated using bond dimension $\chi$ as $\xi_{1}(\chi)$ and $\xi_{2}(\chi)$ respectively; the larger the bond dimension, the more accurate the estimation.
\end{document}